\documentclass[pra,amsfonts,amssymb,nofootinbib]{revtex4}
\usepackage{amsmath}
\usepackage{amssymb}
\usepackage{graphics}
\usepackage{latexsym}

\begin{document}
\bibliographystyle{revtex}

\def\Bid{{\mathchoice {\rm {1\mskip-4.5mu l}} {\rm
{1\mskip-4.5mu l}} {\rm {1\mskip-3.8mu l}} {\rm {1\mskip-4.3mu l}}}}

\newcommand{\eL}{{\cal L}}
\newcommand{\half}{\frac{1}{2}}
\newcommand{\J}{\textbf{J}}
\newcommand{\bP}{\textbf{P}}
\newcommand{\G}{\textbf{G}}
\newcommand{\K}{\textbf{K}}
\newcommand{\M}{{\cal M}}
\newcommand{\E}{{\cal E}}
\newcommand{\bu}{\textbf{u}}
\newcommand{\tr}{\mbox{tr}}
\newcommand{\norm}[1]{\left\Vert#1\right\Vert}
\newcommand{\abs}[1]{\left\vert#1\right\vert}
\newcommand{\set}[1]{\left\{#1\right\}}
\newcommand{\ket}[1]{\left\vert#1\right\rangle}
\newcommand{\bra}[1]{\left\langle#1\right\vert}
\newcommand{\ele}[3]{\left\langle#1\left\vert#2\right\vert#3\right\rangle}
\newcommand{\inn}[2]{\left\langle#1\vert#2\right \rangle}
\newcommand{\Real}{\Bid R}
\newcommand{\dmat}[2]{\ket{#1}\!\!\bra{#2}}

\title{Generalized Euler Angle Parametrization for $SU(N)$}

\author{Todd Tilma}
\affiliation{The Ilya Prigogine Center for Studies in Statistical Mechanics
and Complex Systems \\
Physics Department \\
The University of Texas at Austin \\
Austin, Texas 78712-1081}
\email[Email:]{tilma@physics.utexas.edu}

\author{E.C.G. Sudarshan}
\affiliation{Center for Particle Physics \\
Physics Department \\
The University of Texas at Austin \\
Austin, Texas 78712-1081}
\email[Email:]{sudarshan@physics.utexas.edu}

\date{\today}

\begin{abstract}
In a previous paper \cite{Tilma1} an Euler angle 
parametrization for $SU(4)$ was given.  Here we present the derivation 
of a generalized Euler angle 
parametrization for $SU(N)$.  The formula for the calculation of the Haar
measure for $SU(N)$ as well as its relation to Marinov's volume formula
for $SU(N)$ \cite{Marinov, Marinov2} will
also be derived.  As an example of this parametrization's
usefulness, the density matrix parametrization and invariant volume
element 
for a qubit/qutrit, three qubit, and two three-state systems, 
also known as two qutrit systems \cite{Preskill}, will also be given.
\end{abstract}

\maketitle



\pagebreak

\section{Introduction}

The importance of group theory in understanding quantum mechanical 
processes has grown over the past 100 years as more 
and more physicists recognized that 
\begin{quote}
The basic principles of quantum mechanics seem to require the
postulation of a Lie algebra of observables and a representation of
this algebra by skew-Hermitian operators on a Hilbert
space \cite{Hermann}.
\end{quote}
Today one finds applications of group theory in numerous research
areas; from high-energy particle theory to experimental nano-scale
physics.  Any physical system that exhibits symmetries, or that can be
thought of as Hermann indicated above, will have a group associated
with it.  In this paper we most concerned with representations of, and 
applications on, $SU(N)$.
This group shows up in numerous areas of study, most notably 
in numerical calculations concerning 
entanglement and other quantum information parameters.  In order to
assist in these numerical calculations, we have produced a 
parametrization of $SU(N)$ that should offer some computational benefit.

We will begin this paper by 
deriving a general Euler angle parametrization for $SU(N)$.
Afterward, a general equation for the differential volume element, otherwise known
as the Haar measure, for
$SU(N)$ will be derived.  Then we will show that this
parametrization yields the familiar invariant volume element,
generated from the integration of the Haar measure, for
$SU(N)$ as derived by Marinov \cite{Marinov, Marinov2}.  Finally, the
parametrization of N by N density matrices with regards to the general
Euler angle parametrization for $SU(N)$ will be shown.  As an illustration of
the usefulness of the parametrization,  
representations of qubit/qutrit, three qubit and two qutrit states will be given.

\section{Review: Euler Angle Parametrization From $SU(2)$ To $SU(4)$}

For a $U \in SU(2)$ the Euler angle representation can be found in any good 
textbook on quantum mechanics or Lie algebras (see for example 
\cite{Greiner, Artin, Sattinger})
\begin{equation}
U = e^{i\sigma_3 \alpha_1}e^{i\sigma_2 \alpha_2}e^{i\sigma_3 \alpha_3}.
\end{equation}
For a $U \in SU(3)$ the Euler angle parametrization was initially given in
\cite{MByrd1, MByrdp1} and later in \cite{Gibbons}
\begin{equation}
\label{su3eas}
U = e^{i\lambda_3 \alpha_1}e^{i\lambda_2 \alpha_2}e^{i\lambda_3
\alpha_3}e^{i\lambda_5 \alpha_4}e^{i\lambda_3 \alpha_5}e^{i\lambda_2
\alpha_6}e^{i\lambda_3 \alpha_7}e^{i\lambda_8 \alpha_8}.
\end{equation}
For a $U \in SU(4)$ the Euler angle parametrization was initially given in \cite{Tilma1}
\begin{eqnarray}
\label{eq:su4eas}
U &=& e^{i\lambda_3 \alpha_1}e^{i\lambda_2 \alpha_2}e^{i\lambda_3 \alpha_3}e^{i\lambda_5 \alpha_4}e^{i\lambda_3 \alpha_5}e^{i\lambda_{10} \alpha_6}e^{i\lambda_3 \alpha_7}e^{i\lambda_2 \alpha_8} \nonumber \\
&&\times e^{i\lambda_3 \alpha_{9}}e^{i\lambda_5 \alpha_{10}}e^{i\lambda_3 \alpha_{11}}e^{i\lambda_2 \alpha_{12}}e^{i\lambda_3 \alpha_{13}}e^{i\lambda_8 \alpha_{14}}e^{i\lambda_{15} \alpha_{15}}.
\end{eqnarray}
We would like to extend this work for a $U \in SU(N)$.

\section{Lie Algebra for $SU(N)$}
\label{sec:lasun}

From \cite{Greiner} we already know how to construct the $\lambda_i$ for
arbitrary $SU(N)$.\footnote{Georgi \cite{Georgi} also gives a method for 
constructing the $N^2-1$ $\lambda_i$ matrices for $SU(N)$.}
\begin{enumerate}
\item{For every $i,j = 1,2,3,\ldots,N;\; i < j$, we define two N by N matrices
\begin{equation}
\begin{split}
[\lambda^{\{1\}}(i,j)]_{\mu \nu}&=\delta_{j\mu}\delta_{i\nu}+\delta_{j\nu}\delta_{i\mu},\\
[\lambda^{\{2\}}(i,j)]_{\mu \nu}&=-i(\delta_{i\mu}\delta_{j\nu}-\delta_{i\nu}\delta_{j\mu}),
\end{split}
\end{equation}
which form $N(N-1)$ linearly independent matrices.\footnote{We follow the standard physics
practice by using hermitian generators.}}
\item{Construct a further $N-1$ matrices according to
\begin{equation}
\lambda_{n^2-1}=\sqrt{\frac{2}{n^2-n}}\begin{pmatrix}
(1 & 0 & 0 & 0 &\dots & 0\\
0 & 1 & 0 & 0 &\dots & 0\\
0 & 0 & 1)_{n-1} & 0 & \dots & 0\\
0 & 0 & 0 & -(n-1) & \dots & 0\\
\dots & \dots & \dots & \dots & \dots & \dots \\
0 & 0 & 0 & 0 & \dots & 0\\
\end{pmatrix}_{N \times N}
\end{equation}
for $n=2,3,\ldots,N$.}
\end{enumerate}
By following this convention, $N^2-1$ traceless matrices can be generated.  These
matrices 
form a basis for the corresponding vector space and thus a representation
of the $SU(N)$ generators \cite{Greiner}.  For example, for $N=2$ we 
generate the well-known Euler $\sigma$ matrices ($i < j, \{i,j\} \le 2$):
\begin{align}
\sigma_1=&\,[\lambda^{\{1\}}(1,2)]_{\mu \nu}=\delta_{2\mu}\delta_{1\nu}+\delta_{2\nu}\delta_{1\mu},\nonumber \\
=&\begin{pmatrix}
0 & 1 \\
1 & 0 \\
\end{pmatrix}, \\
\sigma_2=&\,[\lambda^{\{2\}}(1,2)]_{\mu \nu}=-i(\delta_{1\mu}\delta_{2\nu}-\delta_{1\nu}\delta_{2\mu}),\nonumber \\
=&\begin{pmatrix}
0 & -i \\
i & 0\\
\end{pmatrix}, \text{ and } \\
\sigma_3=&\lambda_{2^2-1}=\lambda_3=\sqrt{\frac{2}{2^2-2}}\begin{pmatrix}
1 & 0 \\
0 & -1\\
\end{pmatrix} \nonumber \\
=&\begin{pmatrix}
1 & 0 \\
0 & -1
\end{pmatrix}.
\end{align}

\section{Deriving the Euler Angle Parametrization for $SU(N)$}

By following the work of Biedenharn \cite{Biedenharn} and
Hermann \cite{Hermann} 
we can now generate a Cartan decomposition of $SU(N)$ for $N>2$.\footnote{The 
reason for this will become apparent in the following discussion.}  First, 
we look at the $N$ by $N$, hermitian, traceless, $\lambda_i$ matrices 
formulated in the previous section.
This set is linearly independent and is the lowest dimensional
faithful representation of the $SU(N)$ Lie algebra.  From these matrices we can
then calculate their commutation relations
\begin{gather}
[\lambda_i , \lambda_j] = 2if_{ijk}\lambda_k, \nonumber \\
f_{ijk} = \frac{1}{4i}Tr[[\lambda_i , \lambda_j]\lambda_k], 
\end{gather}
and by observation of the
corresponding structure constants $f_{ijk}$ one can see the
relationship in the algebra that can help generate the Cartan
decomposition of $SU(N)$ (shown for $SU(3)$ in \cite{MByrd1} and for $SU(4)$ in 
\cite{Tilma1}).  

Knowledge of the structure constants allows us to define two subspaces
of the $SU(N)$ group manifold hereafter
known as $K$ and $P$.  
From these subspaces, there corresponds two subsets of the Lie algebra
of $SU(N)$, $L(K)$ and
$L(P)$, such that for $k_1,k_2 \in L(K)$ and $p_1,p_2 \in L(P)$,
\begin{align}
[k_1,k_2] &\in L(K), \nonumber \\
[p_1,p_2] &\in L(K), \nonumber \\
[k_1,p_2] &\in L(P). 
\end{align}
For $SU(N)$, $L(K) = \{ \lambda_{1},\ldots,\lambda_{(N-1)^2-1},\lambda_{N^2-1} \}$  and $L(P) =
\{ \lambda_{(N-1)^2},\ldots,\lambda_{N^2-2} \}$.\footnote{For example, for $SU(4)$ we have 
$L(K)=\{\lambda_{1}, \lambda_{2}, \ldots, \lambda_{8}, \lambda_{15} \}$ and
$L(P)=\{\lambda_{9}, \lambda_{10}, \ldots, \lambda_{14}\}$ \cite{Tilma1}.  
By definition, $L(K)$ and $L(P)$ do not have any elements in common.  
This means that for $N=2$, 
the second $\lambda_1$ element that is generated by this construction as an
element of $L(P)$ must be discarded.  Similarly, the undefined $\lambda_0$ element in $L(K)$
must also be discarded.  For $N \ge 3$ one does not generate any duplications.}
Given that we can
decompose the $SU(N)$ algebra into a semi-direct sum \cite{Herstein}
\begin{equation}
L(SU(N))=L(K) \oplus L(P),
\end{equation}
we therefore have a decomposition of the group,
\begin{equation}
U=K\cdot P.
\end{equation}

From \cite{Sattinger} we know that $L(K)$ contains the generators of the
$SU(N-1)$ subalgebra of $SU(N)$, thus
$K$ will be the $U(N-1)$ subgroup obtained by
exponentiating this subalgebra,
$\{ \lambda_1,\ldots,\lambda_{(N-1)^2-1} \}$, combined with
$\lambda_{N^2-1}$ and thus can be written as (see \cite{MByrd1, Tilma1} for examples)
\begin{equation}
K=[SU(N-1)]\cdot e^{i\lambda_{N^2-1} \phi}
\end{equation}
where $[SU(N-1)]$ represents the $(N-1)^2-1$ term Euler angle 
representation of the $SU(N-1)$ subgroup.

Now, as for $P$, of the $2(N-1)$ elements in $L(P)$ we chose the
$\lambda_2$ analogue, $\lambda_{X_{SU(N)}}$, for $SU(N)$ and write any element of $P$ as
\begin{equation}
P=K' \cdot e^{i\lambda_{X_{SU(N)}}\psi} \cdot K''
\end{equation}
where $K'$ and $K''$ are copies of $K$ and $\lambda_{X_{SU(N)}}$ is given by the $N$ by $N$ matrix
\begin{equation}
\lambda_{X_{SU(N)}}=\lambda_{(N-1)^2+1}=\begin{pmatrix}
0 & 0 & 0 & \dots & 0 & -i\\
0 & 0 & 0 & \dots & 0 & 0\\
0 & 0 & 0 & \dots & 0 & 0\\
\dots & \dots & \dots & \dots & \dots &\dots\\
0 & 0 & 0 & \dots & 0 & 0\\
i & 0 & 0 & \dots & 0 & 0\\
\end{pmatrix}_{N \times N}.
\end{equation}

Unfortunately, at this point in our derivation, 
we have over parameterized $U$ by $2N^2-6N+5$ elements
\begin{equation}
U=K \cdot K' \cdot e^{i\lambda_{(N-1)^2+1}\psi} \cdot K''.
\end{equation}
But, if we recall that $U$ is a product of operators in $SU(N)$, we
can ``remove the redundancies,'' i.e.\ the first $K'$ component as well as the 
$N-1$ Cartan subalgebra elements of $SU(N)$ in the original $K$ component, 
to arrive at the following product (again, see \cite{MByrd1, MByrdp1, Tilma1}
for examples)
\begin{eqnarray}
U &=& [SU(N-1)]\cdot e^{i\lambda_{N^2-1}\phi}e^{-i\lambda_{N^2-1}\phi}e^{-i\lambda_{(N-1)^2-1}\phi_1}
e^{-i\lambda_{(N-2)^2-1}\phi_2}\cdots e^{-i\lambda_{3}\phi_{N-1}} \nonumber \\
&&\times\; e^{i\lambda_{(N-1)^2+1}\psi}\cdot [SU(N-1)]\cdot e^{i\lambda_{N^2-1}\phi} \nonumber \\
&=& [SU(N-1)]\cdot e^{-i\lambda_{(N-1)^2-1}\phi_1}\cdots
e^{-i\lambda_{3}\phi_{N-1}}
e^{i\lambda_{(N-1)^2+1}\psi}\cdot [SU(N-1)]\cdot e^{i\lambda_{N^2-1}\phi}.
\end{eqnarray}
By insisting that our parametrization must truthfully reproduce known vector 
and tensor
transformations under $SU(N)$, we can remove the last ``redundancy,'' the final 
$N^2-5N+5$ elements in $K$,\footnote{For $N=3$, there are no redundancies.  In fact, 
the -1, ($3^2-5*3-5=-1$), that occurs here is a result of removing one too many 
Cartan subalgebra elements from the end of $K$ 
in the previous step.  For $N=3$, one must \textit{restore} a
Cartan subalgebra element, in this case $e^{i\lambda_3\alpha_4}$, back into its
original position in the $K$ component.  For $N>3$ this situation
does not occur.} and, 
after rewriting the parameters, get:
\begin{align}
U=&\biggl(\prod_{2 \leq k \leq N}A(k)\biggr)\cdot [SU(N-1)]\cdot e^{i\lambda_{N^2-1} \alpha_{N^2-1}}, \nonumber\\
A(k)=&\; e^{i\lambda_{3} \alpha_{(2k-3)}}e^{i\lambda_{(k-1)^2+1} \alpha_{2(k-1)}}. \label{eq:suNrec}
\end{align}
This equation, effectively a recurrence
relation for the Euler angle decomposition of $SU(N)$, can be further
rewritten into a more explicit, and therefore final, form
\begin{align}
U=&\biggl(\prod_{2 \leq k \leq N}A(k,j(N))\biggr) \cdot
\biggl(\prod_{2 \leq k \leq N-1}A(k,j((N-1))\biggr) \cdots
\biggl(A(2,j(2))\biggr)\nonumber \\
&\times
e^{i\lambda_{3} \alpha_{N^2-(N-1)}} \cdots e^{i\lambda_{(N-1)^2-1}
  \alpha_{N^2-2}} e^{i\lambda_{N^2-1} \alpha_{N^2-1}} \nonumber\\
=&\prod_{N \geq m \geq 2}\;\biggl(\prod_{2 \leq k \leq m}
A(k,j(m))\biggr) \nonumber \\
&\times e^{i\lambda_{3} \alpha_{N^2-(N-1)}} \cdots e^{i\lambda_{(N-1)^2-1}
  \alpha_{N^2-2}} e^{i\lambda_{N^2-1} \alpha_{N^2-1}},\nonumber \\
A(k,j(m))=&\; e^{i\lambda_{3}
  \alpha_{(2k-3)+j(m)}}e^{i\lambda_{(k-1)^2+1} \alpha_{2(k-1)+j(m)}},
  \nonumber \\
j(m) =&
\begin{cases}
0 \qquad &m=N,\\
\underset{0 \leq l \leq N-m-1}{\sum}2(m+l) \qquad &m \neq N. 
\end{cases}\label{eq:suN}
\end{align}
In this form, the important $\lambda_i$ matrices for equation
(\ref{eq:suN}) are
\begin{align}
\lambda_{3}=&\begin{pmatrix}
1 & 0 & \dots & 0\\
0 & -1 & \dots & 0 \\
\dots & \dots & \dots & \dots \\
0 & 0 & \dots & 0\end{pmatrix}_{N \times N}, \nonumber \\
\lambda_{(k-1)^2+1}=&\begin{pmatrix}
\begin{bmatrix}
0 & \dots & -i \\
\dots & \dots & \dots \\
i & \dots & 0 \end{bmatrix}_{k\times k} & \dots & 0 \\
0 & \dots & 0 \\
\end{pmatrix}_{N \times N} k < N, \nonumber \\
\lambda_{(N-1)^2+1}=&\begin{pmatrix}
0 & \dots & -i \\
\dots & \dots & \dots \\
i & \dots & 0 \\
\end{pmatrix}_{N \times N} k=N, \nonumber \\
\lambda_{N^2-1}=&\sqrt{\frac{2}{N^2-N}}\begin{pmatrix}
1 & 0 & \dots & 0\\
0 & 1 & \dots & 0\\
\dots & \dots & \dots & \dots\\
0 & 0 & \dots & -(N-1)\\
\end{pmatrix}_{N \times N}.
\label{eq:suNmat}
\end{align}
Notice that even though we restricted ourselves to $N>2$ for the 
Cartan decomposition, 
equation (\ref{eq:suN}) is valid for $N\geq2$.  For our purposes it is enough 
to note that this parametrization
is special unitary by construction and can 
be shown to cover the group by modifying the ranges that 
follow and substituting them into the parametrization of 
the characters \cite{Gibbons}.\footnote{  
One may be somewhat distressed by our ``removal of
the redundancies'' statement that precluded the development of 
equation (\ref{eq:suN}) so we offer another, geometrical,
argument for the form of the parametrization in Appendix \ref{app:pictureformalism}.}

\section{Procedure for Calculating the Haar Measure and Group Volume for $SU(N)$}

Taking the Euler angle parametrization given by equation (\ref{eq:suN}) we now
develop the differential volume element, also known as the Haar measure, for the group.  We
proceed by using the method originally given in \cite{Murnaghan} and
developed for $SU(3)$ in \cite{MByrd1, MByrdp1} and for $SU(4)$ in \cite{Tilma1}; 
take a generic $U \in SU(N)$ and find the matrix
\begin{equation}
U^{-1} \cdot dU 
\end{equation}
of left invariant one-forms, then take the determinant
of the matrix of their expansion coefficients.\footnote{A detailed
explanation of this method can be 
found in \cite{Tilma1}.  We should note that one can take 
the determinant of the matrix of the expansion coefficients of the 
$N^2-1$ right invariant one-forms which
also yields the Haar measure in question.  This
is due to the fact that a compact simply-connected real Lie group
has a bi-invariant measure, unique up to a constant factor.  Such a group
is usually referred to as `unimodular' \cite{Sattinger}.} 

To begin, we take the transpose of U to generate
\begin{align}
\label{eq:suNconjeas}
u=&U^T \nonumber\\
=&\; e^{i\lambda_{N^2-1}^T \alpha_{N^2-1}}\cdot[SU(N-1)]^T\cdot(\prod_{N
  \geq k \geq 2}A(k)^T) \nonumber \\
=&\; e^{i\lambda_{N^2-1}^T \alpha_{N^2-1}}e^{i\lambda_{(N-1)^2-1}^T
  \alpha_{N^2-2}} \cdots e^{i\lambda_{3}^T \alpha_{N^2-(N-1)}} \nonumber \\
&\times  \prod_{2 \leq m \leq N}\;\biggl(\prod_{m \geq k \geq
  2}A(k,j(m))^T\biggr)\\ 
\intertext{where}
A(k,j(m))^T=&\; e^{i\lambda_{(k-1)^2+1}^T \alpha_{2(k-1)+j(m)}}e^{i\lambda_{3}^T
  \alpha_{(2k-3)+j(m)}}, \nonumber \\
j(m) =&
\begin{cases}
0 \qquad &m=N,\\
\underset{0 \leq l \leq N-m-1}{\sum}2(m+l) \qquad &m \neq N. 
\end{cases}
\end{align}
An observation of the components of our Lie algebra sub-set 
\begin{eqnarray}
&\{\lambda_{(k-1)^2+1}, \lambda_{k^2-1}\},\nonumber \\
&2 \leq k \leq N,
\end{eqnarray}
shows that the transpose operation is equivalent to making the following substitutions
\begin{align}
\lambda_{(k-1)^2+1}^T &\rightarrow -\lambda_{(k-1)^2+1},\nonumber \\
\lambda_{k^2-1}^T &\rightarrow \lambda_{k^2-1},
\end{align}
for $2 \leq k \leq N$
in equation (\ref{eq:suNconjeas}) generating
\begin{align}
\label{eq:suNconjeasF}
u=&\; e^{i\lambda_{N^2-1} \alpha_{N^2-1}}e^{i\lambda_{(N-1)^2-1}
  \alpha_{N^2-2}} \cdots e^{i\lambda_{3} \alpha_{N^2-(N-1)}} \nonumber \\
&\times \prod_{2 \leq m \leq N}\;\biggl(\prod_{m \geq k \geq
  2}A(k,j(m))'\biggr),\nonumber \\
A(k,j(m))'=&\; e^{-i\lambda_{(k-1)^2+1} \alpha_{2(k-1)+j(m)}}e^{i\lambda_{3}
  \alpha_{(2k-3)+j(m)}}, \nonumber \\
j(m) =&
\begin{cases}
0 \qquad &m=N,\\
\underset{0 \leq l \leq N-m-1}{\sum}2(m+l) \qquad &m \neq N. 
\end{cases}
\end{align}
 
With this form in hand, we then 
take the partial derivative of $u$ with respect to each of the
$N^2-1$ parameters.  In general, the differentiation will have the form
\begin{equation}
\label{partialgen}
\frac{\partial u}{\partial \alpha_{l}} =  E^LC(\alpha_l)E^{-L}u
\end{equation}
where we have used a ``shorthand'' representation with
\begin{eqnarray}
&C(\alpha_l)\in i*\{-\lambda_{(k-1)^2+1}, \lambda_{k^2-1}\}, \nonumber \\
&2 \leq k \leq N,
\end{eqnarray}
and
\begin{align}
E^L=&\; e^{C(\alpha_{N^2-1})
  \alpha_{N^2-1}}\cdots e^{C(\alpha_{l+1})
  \alpha_{l+1}} ,\nonumber\\
E^{-L}=&\; e^{-C(\alpha_{l+1}) \alpha_{l+1}}\cdots e^{-C(\alpha_{N^2-1})
  \alpha_{N^2-1}}.
\end{align} 
By using these equations and the Baker-Campbell-Hausdorff relation,
\begin{equation}
 e^{X}Ye^{-X} = Y + [X,Y] + \frac{1}{2}[X,[X,Y]] + \ldots,
\end{equation}
we are able to consecutively solve equation \eqref{partialgen} for
$l=\{N^2-1,\ldots,1\}$, 
leading to a set of $N$ by $N$ matrices which can be expanded in terms 
of the $N^2-1$ \textit{transposed} elements of the $SU(N)$ Lie algebra with
expansion coefficients $c_{lj}$ given by trigonometric functions of the group parameters $\alpha_i$:
\begin{equation}
M_{l}\equiv \frac{\partial u}{\partial \alpha_{l}} u^{-1} = E^{L}C(\alpha_l)E^{-L} =
\sum_{N^2-1 \ge j \ge 1} c_{lj}\lambda_j^T.
\end{equation}
The $\lambda_j^T$'s can be generated by using the methods contained in section \ref{sec:lasun}.  

Now the coefficients $c_{lj}$'s are the elements of the determinant 
in question.  They are found by evaluating the following trace \cite{Greiner}:
\begin{equation}
c_{lj} = \frac{-i}{2}Tr[\lambda_j^T\cdot M_{l}].
\end{equation}  
The index $l$ corresponds to the specific $\alpha$ parameter, 
the $j$ corresponds to the specific \textit{transposed} element of the algebra.  
Both the $l$ and $j$ indices run from $N^2-1$ to 1.
The determinant of this $N^2-1$ by $N^2-1$ matrix yields the 
differential volume element, also known as the Haar measure for the group, $dV_{SU(N)}$
that, when integrated over the correct values for the ranges of the
parameters and multiplied by a derivable 
normalization constant, yields the volume for the group.

The full $N^2-1$ by $N^2-1$ determinant $\text{Det}[c_{lj}]$, $l,j \in
\{N^2-1,\ldots,1\}$, can be done, or one can notice that the
matrix can be written as
\begin{equation}
C_{SU(N)}= \begin{Vmatrix}
  c_{N^2-1,N^2-2} & c_{N^2-1,N^2-3} &\dots& c_{N^2-1,1} & c_{N^2-1,N^2-1}\\
  c_{N^2-2,N^2-2} & c_{N^2-2,N^2-3} &\dots& c_{N^2-1,1} & c_{N^2-2,N^2-1}\\
  \dots & \dots & \dots & \dots & \dots \\
  c_{1,N^2-2} & c_{1,N^2-3} &\dots& c_{1,1} & c_{1,N^2-1}
\end{Vmatrix}
\end{equation}
which differs only by an overall sign from $\text{Det}[c_{lj}]$ above, but yields a quasi-block form that generates
\begin{equation}
C_{SU(N)} =\begin{Vmatrix}
   O & R \\
   T & Q 
\end{Vmatrix}.
\end{equation}
In this form, $R$ corresponds to the $(N-1)^2$ by $(N-1)^2$ matrix whose determinant is
equivalent to 
$dV_{SU(N-1)}\cdot d\alpha_{N^2-1}$, $Q$ is a complicated
$2(N-1)$ by $(N-1)^2$
matrix, and $O$ is a $(N-1)^2$ by $2(N-1)$ matrix
whose elements are all zero.  Now the interchange of
two columns of a $N$ by $N$ matrix yields a change in sign
of the corresponding determinant.  But by moving $2(N-1)$ columns at once, the
sign of the determinant does not change, and thus one may
generate a new matrix which is now block diagonal
\begin{equation}
C_{SU(N)} =\begin{Vmatrix}
   R & O \\
   Q & T 
\end{Vmatrix}.
\end{equation}
Thus, the full determinant is just equal to the determinant of the
diagonal blocks, one of which is already known.  So only the determinant of the $2(N-1)$ by $2(N-1)$
sub-matrix $T$,
\begin{equation}
\text{Det}[T]= \begin{Vmatrix}
\  c_{2(N-1),N^2-2} &\dots& c_{2(N-1),(N-1)^2}\\
  \dots &\dots& \dots \\
  c_{1,N^2-2} &\dots& c_{1,(N-1)^2}
\end{Vmatrix},
\end{equation}
is needed.  Therefore the Haar measure for $SU(N)$ is nothing more than
\begin{eqnarray}
\label{dvsun}
dV_{SU(N)} &=& \text{Det}[c_{lj}]\nonumber\\ 
&=& -\text{Det}[T]*\text{Det}[D]d\alpha_{N^2-1}\ldots d\alpha_{1}\nonumber\\
&=& -\text{Det}[T]*dV_{SU(N-1)}d\alpha_{N^2-1}d\alpha_{2(N-1)}\ldots d\alpha_{1}.
\end{eqnarray}
This is determined up to normalization (explained in detail in Appendix \ref{app:Haar}).  
Integration over the $N^2-1$ parameter space yields the following  
group volume formula (for $N \geq 2$)
\begin{equation}
\label{suvol}
V_{SU(N)}=\idotsint\limits_V dV_{SU(N)}= \Omega_N*\idotsint\limits_{V^\prime}
dV_{SU(N)}
\end{equation}
where
\begin{eqnarray}
\label{omeganorm}
\Omega_N&=&2^{N-2}*\Omega_{(N-1)}*N \nonumber \\
&=&2^{N-2}*2^{N-3}\cdots 2^0 * 1*2\cdots (N-1)*N \nonumber \\
&=&2^\epsilon N!
\end{eqnarray}
and
\begin{eqnarray}
\epsilon&=&\sum_{2 \leq l \leq N}(N-l) \nonumber \\
&=&(N-2)+(N-3)+(N-4)+ \cdots + (N-(N-1))+(N-N) \nonumber \\
&=&(N-2)+(N-3)+(N-4)+ \cdots + 1 + 0 \nonumber \\
&=&\frac{1}{2}(N-2)(N-1).
\end{eqnarray}

\section{Example: SU(5) Parametrization, Haar Measure, and Group Volume}

Let $U \in SU(5)$. 
Following the methodology developed in the previous section,
the Euler angle parametrization for $SU(5)$ can be seen to be given by
\begin{align}
U=& \prod_{5 \geq m \geq 2}\;\biggl(\prod_{2 \leq k \leq m}
A(k,j(m))\biggr) \nonumber \\
&\times e^{i\lambda_{3} \alpha_{21}}e^{i\lambda_{8}
  \alpha_{22}} e^{i\lambda_{15} \alpha_{23}}e^{i\lambda_{24} \alpha_{24}},
\nonumber \\
A(k,j(m))=&\; e^{i\lambda_{3}
  \alpha_{(2k-3)+j(m)}}e^{i\lambda_{(k-1)^2+1} \alpha_{2(k-1)+j(m)}},
  \nonumber \\
j(m) =&
\begin{cases}
0 \qquad &m=5,\\
\underset{0 \leq l \leq 5-m-1}{\sum}2(m+l) \qquad &m \neq 5. 
\end{cases}\label{eq:su5}
\end{align}
Expansion yields
\begin{align}
U=&\biggl(\prod_{2 \leq k \leq 5}
A(k,j(5))\biggr) \biggl(\prod_{2 \leq k \leq 4}
A(k,j(4))\biggr) \biggl(\prod_{2 \leq k \leq 3}
A(k,j(3))\biggr) \biggl(\prod_{2 \leq k \leq 2}
A(k,j(2))\biggr) \nonumber \\ 
&\times e^{i\lambda_{3} \alpha_{21}}e^{i\lambda_{8}
  \alpha_{22}} e^{i\lambda_{15} \alpha_{23}}e^{i\lambda_{24} \alpha_{24}} 
\nonumber \\
=&A(2,j(5))A(3,j(5))A(4,j(5))A(5,j(5))A(2,j(4))A(3,j(4))A(4,j(4))
A(2,j(3))A(3,j(3))A(2,j(2)) \nonumber \\
&\times e^{i\lambda_{3} \alpha_{21}}e^{i\lambda_{8}
  \alpha_{22}} e^{i\lambda_{15} \alpha_{23}}e^{i\lambda_{24} \alpha_{24}}.
\end{align}
The $j(m)$ values are
\begin{align}
j(5)=&0, \nonumber \\
j(4)=&\underset{0 \leq l \leq 5-4-1}{\sum}2(m+l)=2m=8, \nonumber \\
j(3)=&\underset{0 \leq l \leq 5-3-1}{\sum}2(m+l)=\underset{0 \leq l \leq 1}{\sum}2(m+l)=2m+2(m+1)=14,
\nonumber \\
j(2)=&\underset{0 \leq l \leq 5-2-1}{\sum}2(m+l)=\underset{0 \leq l \leq 2}{\sum}2(m+l)=2m+2(m+1)+2(m+2)
=18,
\end{align}
and the $A(k,j(m))$ components are
\begin{align}
A(2,j(5))=&\; e^{i\lambda_{3}\alpha_{1}}e^{i\lambda_{2} \alpha_{2}}, \nonumber \\
A(3,j(5))=&\; e^{i\lambda_{3}\alpha_{3}}e^{i\lambda_{5} \alpha_{4}}, \nonumber \\
A(4,j(5))=&\; e^{i\lambda_{3}\alpha_{5}}e^{i\lambda_{10} \alpha_{6}}, \nonumber \\
A(5,j(5))=&\; e^{i\lambda_{3}\alpha_{7}}e^{i\lambda_{17} \alpha_{8}}, \nonumber \\
A(2,j(4))=&\; e^{i\lambda_{3}\alpha_{1+8}}e^{i\lambda_{2} 
\alpha_{2+8}}=e^{i\lambda_{3}\alpha_{9}}e^{i\lambda_{2} \alpha_{10}}, \nonumber \\
A(3,j(4))=&\; e^{i\lambda_{3}\alpha_{3+8}}e^{i\lambda_{5} 
\alpha_{4+8}}=e^{i\lambda_{3}\alpha_{11}}e^{i\lambda_{5} \alpha_{12}}, \nonumber \\
A(4,j(4))=&\; e^{i\lambda_{3}\alpha_{5+8}}e^{i\lambda_{10} 
\alpha_{6+8}}=e^{i\lambda_{3}\alpha_{13}}e^{i\lambda_{10} \alpha_{14}}, \nonumber \\
A(2,j(3))=&\; e^{i\lambda_{3}\alpha_{1+14}}e^{i\lambda_{2}
\alpha_{2+14}}=e^{i\lambda_{3}\alpha_{15}}e^{i\lambda_{2} \alpha_{16}}, \nonumber \\
A(3,j(3))=&\; e^{i\lambda_{3}\alpha_{3+14}}e^{i\lambda_{5} 
\alpha_{4+14}}=e^{i\lambda_{3}\alpha_{17}}e^{i\lambda_{5} \alpha_{18}}, \nonumber \\
A(2,j(2))=&\; e^{i\lambda_{3}\alpha_{1+18}}e^{i\lambda_{2} 
\alpha_{2+18}}=e^{i\lambda_{3}\alpha_{19}}e^{i\lambda_{2} \alpha_{20}}.
\end{align}
Thus
\begin{align}
U =&\; e^{i\lambda_3 \alpha_1}e^{i\lambda_2 \alpha_2}e^{i\lambda_3 \alpha_3}e^{i\lambda_5 \alpha_4}e^{i\lambda_3 \alpha_5}e^{i\lambda_{10} \alpha_6}e^{i\lambda_3 \alpha_7}e^{i\lambda_{17} \alpha_8}
e^{i\lambda_3 \alpha_9}e^{i\lambda_2 \alpha_{10}}e^{i\lambda_3 \alpha_{11}}e^{i\lambda_5 \alpha_{12}}e^{i\lambda_3 \alpha_{13}}e^{i\lambda_{10} \alpha_{14}} \nonumber \\
&\times e^{i\lambda_3 \alpha_{15}}e^{i\lambda_2 \alpha_{16}}e^{i\lambda_3 \alpha_{17}}e^{i\lambda_5 \alpha_{18}}e^{i\lambda_3 \alpha_{19}}e^{i\lambda_2 \alpha_{20}}e^{i\lambda_3 \alpha_{21}}
e^{i\lambda_8 \alpha_{22}}e^{i\lambda_{15} \alpha_{23}}e^{i\lambda_{24} \alpha_{24}}, 
\label{eq:su5eas}
\end{align}
with a differential volume element of
\begin{align}
dV_{SU(5)} =& \text{Det}[T]dV_{SU(4)}d\alpha_{24}d\alpha_{8}\ldots d\alpha_{1}
\nonumber \\
=&\text{Det}[T]\cos(\alpha_{12})^3\cos(\alpha_{14})\cos(\alpha_{18})\sin(2\alpha_{10})\sin(\alpha_{12}) \nonumber \\
&\times \sin(\alpha_{14})^5\sin(2\alpha_{16})\sin(\alpha_{18})^3\sin(2\alpha_{20})d\alpha_{24}\ldots d\alpha_{1},
\end{align}
and where $\text{Det}[T]$ is an 8 by 8 matrix composed of the following elements
\begin{align}
c_{lj} =& \frac{-i}{2}Tr[\lambda_j \cdot M_{l}], \nonumber \\
 =& \frac{-i}{2}Tr[\lambda_j \cdot \biggr(\frac{\partial u}{\partial \alpha_{l}} u^{-1}\biggl)] \qquad 
(1 \leq l \leq 8, \; 16 \leq j \leq 23,\; u=U^T),
\end{align}
which when calculated yields
\begin{equation}
\text{Det}[T]=\cos(\alpha_{4})^3\cos(\alpha_{6})^5\cos(\alpha_{8})\sin(2\alpha_{2})
\sin(\alpha_{4})\sin(\alpha_{6})\sin(\alpha_{8})^7.
\end{equation}
Explicit calculation of the invariant volume element for $SU(5)$ can be done by
using equations \eqref{suvol}, \eqref{omeganorm} and 
the material from Appendix \ref{app:Haar}.  From this
one generates
\begin{align}
V_{SU(5)}=&\;  2^6 * 5! * V(SU(5)/Z_5) \nonumber\\
=&\frac{\sqrt{5}\pi^{14}}{72},
\end{align}
which is in agreement with 
the volume obtained by Marinov \cite{Marinov2}.

It should be noted that in calculating $V_{SU(5)}$ 
one must use the following 
ranges of integration from Appendix \ref{app:Haar} and 
expressed in equation \eqref{suvol}
as $\text{V}'$:
\begin{alignat}{2}
0 \le \alpha_{2i-1} \le \pi, &\qquad 0 \le \alpha_{2i} \le
\frac{\pi}{2},\quad (1 \leq i \leq 10)\nonumber \\
0 \le \alpha_{21} \le \pi, &\qquad 0 \le \alpha_{22} \le
\frac{\pi}{\sqrt{3}}, \nonumber \\
0 \le \alpha_{23} \le \frac{\pi}{\sqrt{6}}, &\qquad 0 \le \alpha_{24}
\le \frac{\pi}{\sqrt{10}}.
\end{alignat}
Note that these ranges \textit{do not} cover the
group $SU(5)$, but rather $SU(5)/Z_5$.  The covering ranges for $SU(5)$,
following the work in Appendix \ref{app:paramranges}, are as follows
\begin{gather}
0 \le \alpha_1, \alpha_9,\alpha_{15},\alpha_{19} \le \pi,\nonumber\\
0 \le \alpha_2,\alpha_4,\alpha_6,\alpha_8,\alpha_{10},\alpha_{12},
\alpha_{14},\alpha_{16},\alpha_{18},\alpha_{20}
\le \frac{\pi}{2},\nonumber\\
0 \le
\alpha_3,\alpha_5,\alpha_7,\alpha_{11},\alpha_{13},\alpha_{17},
\alpha_{21} \le 2\pi,\nonumber\\
0 \le \alpha_{22} \le \sqrt{3}\pi,\nonumber\\
0 \le \alpha_{23} \le 2\sqrt{\frac{2}{3}}\pi,\nonumber\\
0 \le \alpha_{24} \le \sqrt{\frac{5}{2}}\pi.
\end{gather}

\section{Generalized Differential Volume Element (Haar Measure) for $SU(N)$}

Let us now quickly review the differential volume element kernels for the first
few $SU(N)$ groups (here rewritten in the order of the parametrization
of the group):
\begin{equation}
\label{su2dv}
dV_{SU(2)}=\sin(2\alpha_{2})d\alpha_{3}d\alpha_{2}d\alpha_{1},
\end{equation}
\begin{align}
dV_{SU(3)}=&\sin(2\alpha_{2})\cos(\alpha_{4})\sin(\alpha_{4})^3\times
dV_{SU(2)} d\alpha_{8}d\alpha_{4}\ldots d\alpha_{1}\nonumber \\
=&\sin(2\alpha_{2})\cos(\alpha_{4})\sin(\alpha_{4})^3\sin(2\alpha_{6})d\alpha_{8}
\ldots d\alpha_{1}, 
\label{eq:su3dv}
\end{align}
\begin{align}
dV_{SU(4)}=&\sin(2\alpha_{2})\cos(\alpha_{4})^3\sin(\alpha_{4})\cos(\alpha_{6})
\sin(\alpha_{6})^5\times dV_{SU(3)}d\alpha_{15}d\alpha_{6}\ldots d\alpha_{1} \nonumber \\
=&
\sin(2\alpha_{2})\cos(\alpha_{4})^3\sin(\alpha_{4})\cos(\alpha_{6})\sin(\alpha_{6})^5
\sin(2\alpha_{8})\cos(\alpha_{10})\sin(\alpha_{10})^3\sin(2\alpha_{12})d\alpha_{15}\ldots
d\alpha_{1},
\label{eq:su4dv}
\end{align}
\begin{align}
dV_{SU(5)}=&
\sin(2\alpha_{2})\cos(\alpha_{4})^3\sin(\alpha_{4})\cos(\alpha_{6})^5\sin(\alpha_{6})\cos(\alpha_{8})\sin(\alpha_{8})^7
\times dV_{SU(4)} d\alpha_{24} d\alpha_{8} \ldots d\alpha_{1} \nonumber \\
=&\sin(2\alpha_{2})\cos(\alpha_{4})^3\sin(\alpha_{4})\cos(\alpha_{6})^5\sin(\alpha_{6})\cos(\alpha_{8})\sin(\alpha_{8})^7
\nonumber \\
&\times \sin(2\alpha_{10})\cos(\alpha_{12})^3\sin(\alpha_{12})\cos(\alpha_{14})\sin(\alpha_{14})^5
\sin(2\alpha_{16})\cos(\alpha_{18})\sin(\alpha_{18})^3\sin(2\alpha_{20})d\alpha_{24}
\ldots d\alpha_{1}.
\label{eq:su5dv}
\end{align}
A pattern is emerging with regard to the trigonometric
components of the differential volume element.  For example, when
one looks at the parametrization of the group and matches the
trigonometric function in the differential volume element kernel with
its corresponding $e^{i\lambda_2\alpha_m}$ component, one sees $\sin(2\alpha_m)$
terms showing up.  It is plain to see then that, in general, the differential 
volume kernel is made up of trigonometric functions
that correspond to group elements that are of the form
$e^{i\lambda_{(k-1)^2+1} \alpha_{2(k-1)+j(m)}}$ and where $j(m)$ is
given in equation (\ref{eq:suN}).  Therefore, there should be a
general expression for the differential volume element kernel for
$SU(N)$ that can be written down via inspection of the Euler angle
parametrization.  We shall now show that this is indeed true, and
give the methodology for writing down the differential volume element
kernel for $SU(N)$.  We will also show that this procedure, after
integration, yields Marinov's volume formula for $SU(N)$.

To begin, let us take a look at the differential volume elements and
their corresponding parametrizations for
$SU(3)$ and $SU(4)$.  From the parametrization originally given in
\cite{MByrd1, MByrdp1} for $SU(3)$ we see that the 
$e^{i\lambda_{5}\alpha_{4}}$ term contributes to the 
$\cos(\alpha_{4})\sin(\alpha_{4})^3$ term in equation 
(\ref{eq:su3dv}).  Yet in the parametrization for $SU(4)$, originally
given in \cite{Tilma1}, the first $e^{i\lambda_{5}\alpha_{4}}$ term yields 
\textit{not} $\cos(\alpha_{4})\sin(\alpha_{4})^3$ but rather 
$\cos(\alpha_{4})^3\sin(\alpha_{4})$.  It is the second installment of
$\lambda_{5}$ in the parametrization for $SU(4)$, 
which can be seen to occur because of equation
(\ref{eq:suNrec}), which gives us the $\cos(\alpha_{4})\sin(\alpha_{4})^3$
term.  For completeness, we should note that the
$\cos(\alpha_{6})\sin(\alpha_{6})^5$ term in the differential volume
element for $SU(4)$ comes from the $e^{i\lambda_{10}\alpha_{6}}$ term in
the parametrization.  

When we now look at the differential volume element and corresponding 
parametrization for $SU(5)$, we see the following relationships
\begin{align}
e^{i\lambda_{5}\alpha_{4}} &\Longrightarrow
\cos(\alpha_{4})^3\sin(\alpha_{4}), \nonumber \\
e^{i\lambda_{10}\alpha_{6}} &\Longrightarrow 
\cos(\alpha_{6})^5\sin(\alpha_{6}), \nonumber \\
e^{i\lambda_{17}\alpha_{8}} &\Longrightarrow 
\cos(\alpha_{8})\sin(\alpha_{8})^7, \nonumber \\
e^{i\lambda_{5}\alpha_{12}} &\Longrightarrow 
\cos(\alpha_{12})^3\sin(\alpha_{12}), \nonumber \\
e^{i\lambda_{10}\alpha_{14}} &\Longrightarrow 
\cos(\alpha_{14})\sin(\alpha_{14})^5, \nonumber \\
e^{i\lambda_{5}\alpha_{18}} &\Longrightarrow 
\cos(\alpha_{18})\sin(\alpha_{18})^3.
\end{align}
By combining all these observations from the parametrizations and 
differential volume element kernels for $SU(2)$ to $SU(5)$ we can see that 
the following pattern is evident 
\begin{align}
e^{i\lambda_2\alpha_2} &\Longrightarrow \sin(2\alpha_2) \quad &k=2, \nonumber \\
e^{i\lambda_{(k-1)^2+1}\alpha_{2(k-1)}} &\Longrightarrow
\cos(\alpha_{2(k-1)})^{2k-3}\sin(\alpha_{2(k-1)}) \quad &2<k<N, \nonumber \\
e^{i\lambda_{(N-1)^2+1}\alpha_{2(N-1)}} &\Longrightarrow
\cos(\alpha_{2(N-1)})\sin(\alpha_{2(N-1)})^{2N-3} \quad &k=N,
\label{eq:sunmethod}
\end{align}
relating the $A(k)$ term of the recurrence relation given in equation
(\ref{eq:suNrec}) with the $\text{Det}[T]$ term of the differential volume
element given in equation \eqref{dvsun}.  Exploiting this recurrence 
relation we now have a methodology for writing down the kernel for 
any $SU(N)$ differential volume element.  Recalling equation
(\ref{eq:suN}) and the $A(k,j(m))$ term we see that in general
\begin{align}
e^{i\lambda_{(k-1)^2+1} \alpha_{2(k-1)+j(m)}}
&\Longrightarrow \sin(2\alpha_{2+j(m)}) &k=2, \nonumber \\
&\Longrightarrow \cos(\alpha_{2(k-1)+j(m)})^{2k-3}\sin(\alpha_{2(k-1)+j(m)}) &2<k<m, 
\nonumber \\
&\Longrightarrow \cos(\alpha_{2(m-1)+j(m)})\sin(\alpha_{2(m-1)+j(m)})^{2m-3} &k=m,
\end{align}
for $N \ge m \ge 2$.  Using this result, combined with the knowledge that
only these parameters contribute to the integrated kernel in
$dV_{SU(N)}$, we are able to write the following product relation for the
kernel of the Haar measure for $SU(N)$
\begin{equation}
\label{dvKsun}
dV_{SU(N)}=K_{SU(N)}d\alpha_{N^2-1}\ldots d\alpha_{1}
\end{equation}
where
\begin{align}
K_{SU(N)}&=\prod_{N \geq m \geq 2}\;\biggl(\prod_{2 \leq k \leq
  m}Ker(k,j(m))\biggr), \nonumber \\
Ker(k,j(m))&=
\begin{cases}
\sin(2\alpha_{2+j(m)}) \quad &k=2, \\
\cos(\alpha_{2(k-1)+j(m)})^{2k-3}\sin(\alpha_{2(k-1)+j(m)}) \quad
&2<k<m, \\
\cos(\alpha_{2(m-1)+j(m)})\sin(\alpha_{2(m-1)+j(m)})^{2m-3} \quad
&k=m,
\end{cases}
\label{eq:dvSUN}
\end{align}
and $j(m)$ is from equation (\ref{eq:suN}).  

\section{Example: SU(6) Haar Measure and Volume Calculation}
\label{sec:su6hm}

As proof of the validity of equation (\ref{eq:dvSUN}) we shall use it to write down the
differential volume element for $SU(6)$.  Observation of equation 
\eqref{dvsun} tells us that
$dV_{SU(6)}$ is dependent on the differential volume elements of 
$SU(5)$, $SU(4)$, $SU(3)$, and 
$SU(2)$.  Thus, in the process of writing down $dV_{SU(6)}$ we will not only confirm the 
calculated differential volume elements for the previous four $SU(N)$ ($N=5,4,3,2$) 
groups, but we will also 
be able to give the functional form of $\text{Det}[T]$ in equation 
\eqref{dvsun} for $SU(6)$ without
having to formally calculate the 10 by 10 determinant.  So, 
for $N=6$, equations (\ref{eq:dvSUN})
and (\ref{eq:suN}) yield
\begin{align}  
K_{SU(6)}&=\prod_{6 \geq m \geq 2}\;\biggl(\prod_{2 \leq k \leq
  m}Ker(k,j(m))\biggr), \nonumber \\
Ker(k,j(m))&=
\begin{cases}
\sin(2\alpha_{2+j(m)}) \quad &k=2, \\
\cos(\alpha_{2(k-1)+j(m)})^{2k-3}\sin(\alpha_{2(k-1)+j(m)}) \quad
&2<k<m, \\
\cos(\alpha_{2(m-1)+j(m)})\sin(\alpha_{2(m-1)+j(m)})^{2m-3} \quad
&k=m,
\end{cases} \nonumber \\
j(m)&=
\begin{cases}
0 \qquad &m=6,\\
\underset{0 \leq l \leq 6-m-1}{\sum}2(m+l) \qquad &m \neq 6. 
\end{cases}
\end{align}
Which when expand gives
\begin{align}
K_{SU(6)}&=\biggl(\prod_{2 \leq k \leq 6}Ker(k,j(6))\biggr)\;\biggl(\prod_{2 \leq k \leq
  5}Ker(k,j(5))\biggr)\;\biggl(\prod_{2 \leq k \leq 4}Ker(k,j(4))\biggr) \nonumber \\
&\times \biggl(\prod_{2 \leq k \leq 3}Ker(k,j(3))\biggr)\;\biggl(\prod_{2 \leq k \leq
  2}Ker(k,j(2))\biggr) \nonumber \\
&=(Ker(2,j(6))Ker(3,j(6))Ker(4,j(6))Ker(5,j(6))Ker(6,j(6))) \nonumber \\
&\times (Ker(2,j(5))Ker(3,j(5))Ker(4,j(5))Ker(5,j(5))) \nonumber \\
&\times (Ker(2,j(4))Ker(3,j(4))Ker(4,j(4))) \nonumber \\
&\times (Ker(2,j(3))Ker(3,j(3))) \nonumber \\
&\times (Ker(2,j(2)).
\end{align}
The $j(m)$ values are
\begin{align}
j(6)=&0, \nonumber \\
j(5)=&\underset{0 \leq l \leq 6-5-1}{\sum}2(m+l)=2m=10, \nonumber \\
j(4)=&\underset{0 \leq l \leq 6-4-1}{\sum}2(m+l)=\underset{0 \leq l \leq 1}{\sum}2(m+l)
=2m+2(m+1)=18,
\nonumber \\
j(3)=&\underset{0 \leq l \leq 6-3-1}{\sum}2(m+l)=\underset{0 \leq l \leq 2}{\sum}2(m+l)
=2m+2(m+1)+2(m+2)=24, \nonumber \\
j(2)=&\underset{0 \leq l \leq 6-2-1}{\sum}2(m+l)=\underset{0 \leq l \leq 3}{\sum}2(m+l)
=2m+2(m+1)+2(m+2)+2(m+3)=28,
\end{align}
and the $Ker(k,j(m))$ components are
\begin{align}
Ker(2,j(6))=&\sin(2\alpha_2), \nonumber \\
Ker(3,j(6))=&\cos(\alpha_4)^3\sin(\alpha_4), \nonumber \\
Ker(4,j(6))=&\cos(\alpha_6)^5\sin(\alpha_6), \nonumber \\
Ker(5,j(6))=&\cos(\alpha_8)^7\sin(\alpha_8), \nonumber \\
Ker(6,j(6))=&\cos(\alpha_{10})\sin(\alpha_{10})^9, \nonumber \\
Ker(2,j(5))=&\sin(2\alpha_{2+10})=\sin(2\alpha_{12}), \nonumber \\
Ker(3,j(5))=&\cos(\alpha_{4+10})^3\sin(\alpha_{4+10})=
\cos(\alpha_{14})^3\sin(\alpha_{14}), \nonumber \\
Ker(4,j(5))=&\cos(\alpha_{6+10})^5\sin(\alpha_{6+10})=
\cos(\alpha_{16})^5\sin(\alpha_{16}), \nonumber \\
Ker(5,j(5))=&\cos(\alpha_{8+10})\sin(\alpha_{8+10})^7=
\cos(\alpha_{18})\sin(\alpha_{18})^7, \nonumber \\
Ker(2,j(4))=&\sin(2\alpha_{2+18})=\sin(2\alpha_{20}), \nonumber \\
Ker(3,j(4))=&\cos(\alpha_{4+18})^3\sin(\alpha_{4+18})=
\cos(\alpha_{22})^3\sin(\alpha_{22}), \nonumber \\
Ker(4,j(4))=&\cos(\alpha_{6+18})\sin(\alpha_{6+18})^5=
\cos(\alpha_{24})\sin(\alpha_{24})^5, \nonumber \\
Ker(2,j(3))=&\sin(2\alpha_{2+24})=\sin(2\alpha_{26}), \nonumber \\
Ker(3,j(3))=&\cos(\alpha_{4+24})\sin(\alpha_{4+24})^3=
\cos(\alpha_{28})\sin(\alpha_{28})^3, \nonumber \\
Ker(2,j(2))=&\sin(2\alpha_{2+28})=\sin(2\alpha_{30}).
\end{align}
Thus
\begin{align}
K_{SU(6)}=&\sin(2\alpha_2)\cos(\alpha_4)^3\sin(\alpha_4)\cos(\alpha_6)^5\sin(\alpha_6)
\cos(\alpha_8)^7\sin(\alpha_8)\cos(\alpha_{10})\sin(\alpha_{10})^9 \nonumber \\
&\times \sin(2\alpha_{12})\cos(\alpha_{14})^3\sin(\alpha_{14})\cos(\alpha_{16})^5
\sin(\alpha_{16})\cos(\alpha_{18})\sin(\alpha_{18})^7 \nonumber \\
&\times \sin(2\alpha_{20})\cos(\alpha_{22})^3\sin(\alpha_{22})\cos(\alpha_{24})
\sin(\alpha_{24})^5 \nonumber \\
&\times \sin(2\alpha_{26})\cos(\alpha_{28})\sin(\alpha_{28})^3 \nonumber \\
&\times \sin(2\alpha_{30})
\end{align}
and
\begin{equation}
dV_{SU(6)}=K_{SU(6)}d\alpha_{35}\ldots d\alpha_1.
\end{equation}
Comparison of the above kernel with those from equations 
\eqref{su2dv}, (\ref{eq:su3dv}),
(\ref{eq:su4dv}), and (\ref{eq:su5dv}) confirms that 
equation (\ref{eq:dvSUN}) does
correctly yield the differential volume element for $SU(6)$.  
As added proof, integration
of $K_{SU(6)}$ using equations \eqref{suvol} and \eqref{omeganorm} 
combined with 
the following ranges (the general derivation of which can be found
in Appendix \ref{app:Haar})
\begin{alignat}{2}
0 \le \alpha_{2i-1} \le \pi, &\qquad 0 \le \alpha_{2i} \le
\frac{\pi}{2},\quad (1 \leq i \leq 15)\nonumber \\
0 \le \alpha_{31} \le \pi, &\qquad 0 \le \alpha_{32} \le
\frac{\pi}{\sqrt{3}}, \nonumber \\
0 \le \alpha_{33} \le \frac{\pi}{\sqrt{6}}, &\qquad 0 \le \alpha_{34}
\le \frac{\pi}{\sqrt{10}}, \nonumber \\
0 \le \alpha_{35} \le \frac{\pi}{\sqrt{15}}, &\qquad
\end{alignat}
yields
\begin{align}
V_{SU(6)}=&2^{10}*6!*V(SU(6)/Z_6) \nonumber \\
=& \frac{\pi^{20}}{1440\sqrt{3}}
\label{eq:su6hmv}
\end{align}
which is in agreement with the volume obtained by Marinov \cite{Marinov2}.

\section{Generalized Group Volume for $SU(N)$}

In looking at equations \eqref{suvol} and (\ref{eq:dvSUN}) we see that under
this Euler angle parametrization for $SU(N)$ the calculation of the invariant group volume is simply
a matter of successive integrations of sines and cosines, multiplied by some
power of $\pi$ and a normalization constant.  Therefore it stands to reason that with 
the derivation of a generalized form for the differential volume element of $SU(N)$,
i.\ e.\ the Haar measure, for $SU(N)$, there should be a corresponding
generalized form for the volume element for $SU(N)$.  It is to this derivation that we now focus our attention.

We begin by noticing that in our Euler angle parametrization we have a total of $N^2-1$
parameters, of which the final $N-1$ are the Cartan subalgebra elements for $SU(N)$, 
thus leaving $N(N-1)$ elements evenly 
split between the $\lambda_{(k-1)^2+1}$ and $\lambda_3$
parameters (for $2\le k \le N$).  Rewriting this observation 
using equation (\ref{eq:suN}) we see
\begin{align}
\frac{N(N-1)}{2} &\Longrightarrow e^{i\lambda_{(k-1)^2+1}\alpha_{2(k-1)+j(m)}}, \nonumber \\
\frac{N(N-1)}{2} &\Longrightarrow e^{i\lambda_3\alpha_{(2k-3)+j(m)}},  \nonumber \\
N-1 &\Longrightarrow e^{i\lambda_3\alpha{N^2-(N-1)}}\ldots e^{i\lambda_{N^2-1}\alpha_{N^2-1}}.
\end{align}
Examination of equations \eqref{dvKsun} and (\ref{eq:dvSUN}) shows that the $N-1$ Cartan
subalgebra elements and the $N(N-1)/2$ $\lambda_3$ elements do not contribute
to the integrated kernel, but their corresponding parameters are 
integrated over.  
Expanding on the general results from Appendix \ref{app:Haar}, we see that we can use 
the following ranges to calculate the group volume for $SU(N)$
\begin{equation}
0 \le \alpha_{2i-1} \le \pi, \quad 0 \le \alpha_{2i} \le \frac{\pi}{2}, \qquad 1 \le i \le 
\frac{N(N-1)}{2},
\end{equation}
and
\begin{equation}
0 \le \alpha_{N^2+b} \le \pi\sqrt{\frac{2}{a(a-1)}}, \qquad a \equiv b\text{ mod }(N+1) \quad a=2
\ldots N.
\end{equation}
From these ranges, it becomes apparent 
that each $\lambda_3$ element contributes a factor of $\pi$ to the total
integration of the differential volume element over the $N^2-1$ parameter space while
each of the $N-1$ Cartan subalgebra elements contributes not only a factor of $\pi$ but
a multiplicative constant as well.  Explicitly
\begin{align}
\int_0^\pi d\alpha_{2i-1} &=\; \pi
\qquad 1\le i \le \frac{N(N-1)}{2} \nonumber \\
\Longrightarrow\; &\pi^\frac{N(N-1)}{2} \quad \text{from the }\lambda_3\text{ components,}
\label{eq:part1V}
\end{align}
and
\begin{align}
\int_0^{\pi\sqrt{\frac{2}{a(a-1)}}}d\alpha_{N^2+b}&=\; \pi\sqrt{\frac{2}{a(a-1)}} \qquad 
a \equiv b\text{ mod }(N+1) \quad a=2\ldots N \nonumber \\
&=\; \pi\sqrt{\frac{2}{k(k-1)}} \qquad 2\le k \le N \nonumber \\
\Longrightarrow\; &\prod_{2\le k \le N}\pi\sqrt{\frac{2}{k(k-1)}} \quad \text{from the N-1 
Cartan subalgebra components.}
\label{eq:part2V}
\end{align}

We now focus our attention on the integration of the differential volume element kernel
given in equation (\ref{eq:dvSUN}).  Examination of the $Ker(k,j(m))$ term combined with the
previously given ranges, yields the following three integrals to be evaluated
\begin{align}
&\int_0^{\frac{\pi}{2}}\sin(2\alpha_{2+j(m)})d\alpha_{2+j(m)} \quad &k=2, \nonumber \\
&\int_0^{\frac{\pi}{2}}\cos(\alpha_{2(k-1)+j(m)})^{2k-3}\sin(\alpha_{2(k-1)+j(m)}))
d\alpha_{2(k-1)+j(m)} \quad
&2<k<m, \nonumber \\
&\int_0^{\frac{\pi}{2}}\cos(\alpha_{2(m-1)+j(m)})\sin(\alpha_{2(m-1)+j(m)})^{2m-3}
)d\alpha_{2(m-1)+j(m)} \quad
&k=m,
\end{align}
and where, again, $j(m)$ is from equation (\ref{eq:suN}).  The first integral is equal to 1,
and the other two can be solved by using the following integral solution from Dwight's ``Tables of
Integrals and Other Mathematical Data''
\begin{equation}
\int_0^{\frac{\pi}{2}}\sin^p(x)\cos^q(x)dx = \frac{\Gamma(\frac{p+1}{2})\Gamma(\frac{q+1}{2})}{2\Gamma(\frac{p+q}{2}+1)} \qquad p+1\;,q+1\;>0
\end{equation}
where $\Gamma(x)$ is the standard Gamma function with the following properties
\begin{align}
\Gamma(1)&= 1, \nonumber \\
\Gamma(x+1)&=x\Gamma(x).
\end{align}
Thus when $p=1$ and $q=2k-3$ we get
\begin{align}
\frac{\Gamma(\frac{1+1}{2})\Gamma(\frac{2k-3+1}{2})}{2\Gamma(\frac{1+2k-3}{2}+1)} &=
\frac{\Gamma(1)\Gamma(k-1)}{2\Gamma(k)} \nonumber \\
&=\frac{\Gamma(k-1)}{2(k-1)\Gamma(k-1)} \nonumber \\
&=\frac{1}{2(k-1)},
\end{align}
and when $p=2m-3$ and $q=1$ we get
\begin{align}
\frac{\Gamma(\frac{2m-3+1}{2})\Gamma(\frac{1+1}{2})}{2\Gamma(\frac{2m-3+1}{2}+1)} &=
\frac{\Gamma(m-1)\Gamma(1)}{2\Gamma(m)} \nonumber \\
&=\frac{\Gamma(m-1)}{2(m-1)\Gamma(m-1)} \nonumber \\
&=\frac{1}{2(m-1)}.
\end{align}

From this work we can see that with regards to the integration of the $Ker(k,j(m))$ terms
in equation (\ref{eq:dvSUN}) their contribution to the overall calculated volume element is
\begin{gather}
\int_0^{\frac{\pi}{2}}\prod_{N \geq m \geq 2}\;\biggl(\prod_{2 \leq k
\leq m}Ker(k,j(m))\biggr)d\alpha_{2i}=\prod_{N \geq m \geq 2}\;
\biggl(\prod_{2 \leq k \leq
m}\mathbb{V}(k,m)\biggr), \nonumber \\
d\alpha_{2i} \text{ runs from } 1\le i \le \frac{N(N-1)}{2},
\end{gather}
and where
\begin{gather}
\mathbb{V}(k,m)=
\begin{cases}
1 \quad &k=2, \\
\frac{1}{2(k-1)} \quad &2<k\le m. 
\end{cases}
\label{eq:vkm}
\end{gather}
Substituting this result as well as equations (\ref{eq:part1V}), (\ref{eq:part2V}), and 
\eqref{dvKsun} into equation \eqref{suvol} yields
\begin{align}
V_{SU(N)}=&\idotsint\limits_V dV_{SU(N)}\nonumber \\
=&\; \Omega_N*\idotsint\limits_{V^\prime}
K_{SU(N)}d\alpha_{N^2-1}\ldots d\alpha_{1} \nonumber \\
=&\; 2^{\frac{(N-1)(N-2)}{2}} N!\; \pi^\frac{N(N-1)}{2}\biggl(\prod_{2\le k \le N}\pi\sqrt{\frac{2}{k(k-1)}}\biggr)
\biggl(\prod_{N \geq m \geq 2}\;\biggl(\prod_{2 \leq k \leq m}\mathbb{V}(k,m)\biggr)\biggr).
\label{eq:vsunUN}
\end{align}
This expression can be simplified by observing that
\begin{align}
\prod_{2\le k \le N}\pi\sqrt{\frac{2}{k(k-1)}}=&\; 2^{\frac{N-1}{2}}\pi^{(N-1)}
\prod_{2\le k \le N}\sqrt{\frac{1}{k(k-1)}} \nonumber \\
=&\; 2^{\frac{N-1}{2}}\pi^{(N-1)}\biggl(\frac{1}{\sqrt{2*1}}*\frac{1}{\sqrt{3*2}}*
\frac{1}{\sqrt{4*3}}*\cdots*\frac{1}{\sqrt{N*(N-1)}}\biggr) \nonumber \\
=&\; 2^{\frac{N-1}{2}}\pi^{(N-1)}\frac{1}{\sqrt{N!(N-1)!}},
\end{align}
and, through the usage of equation (\ref{eq:vkm}), that
\begin{align}
\prod_{N \geq m \geq 2}\;\biggl(\prod_{2 \leq k \leq
m}\mathbb{V}(k,m)\biggr)=&\; \biggl(\prod_{2 \leq k \leq
N}\mathbb{V}(k,N)\biggr)\biggl(\prod_{2 \leq k \leq
N-1}\mathbb{V}(k,N-1)\biggr) \nonumber \\
& \times \cdots \times \biggl(\prod_{2 \leq k \leq
4}\mathbb{V}(k,4)\biggr)\biggl(\prod_{2 \leq k \leq
3}\mathbb{V}(k,3)\biggr)\mathbb{V}(2,2) \nonumber \\
=&(\mathbb{V}(2,N)\mathbb{V}(3,N)\cdots \mathbb{V}(N,N))
(\mathbb{V}(2,N-1)\mathbb{V}(3,N-1)\cdots \mathbb{V}(N-1,N-1)) \nonumber \\
&\times \cdots \times (\mathbb{V}(2,4)\mathbb{V}(3,4)\mathbb{V}(4,4))(\mathbb{V}(2,3)
\mathbb{V}(3,3))\mathbb{V}(2,2)) \nonumber \\
=&\mathbb{V}(3,N)\mathbb{V}(3,N-1)\cdots \mathbb{V}(3,4)\mathbb{V}(3,3) \times 
\mathbb{V}(4,N)\mathbb{V}(4,N-1)\cdots \mathbb{V}(4,4) \nonumber \\
&\times \mathbb{V}(N-1,N)\mathbb{V}(N-1,N-1) \times \mathbb{V}(N,N) \nonumber \\
=&\biggl(\frac{1}{4}\biggr)^{N-2}\times \biggl(\frac{1}{6}\biggr)^{N-3} \times 
\cdots \times \biggl(\frac{1}{2(N-2)}\biggr)^2 \times \biggl(\frac{1}{2(N-1)}\biggr) 
\nonumber \\
=&\biggl(\frac{1}{2}\biggr)^{N-2}\biggl(\frac{1}{2}\biggr)^{N-2}\times  \biggl(\frac{1}{2}\biggr)^{N-3}
\biggl(\frac{1}{3}\biggr)^{N-3} \times \cdots \times \biggl(\frac{1}{2}\biggr)^2 
\biggl(\frac{1}{(N-2)}\biggr)^2 \times \biggl(\frac{1}{2}\biggr)\biggl(\frac{1}{(N-1)}\biggr) 
\nonumber \\
=&\;2^{-\frac{(N-2)(N-1)}{2}}\prod^{N-1}_{k=1}
\biggl(\frac{1}{k!}\biggr).
\end{align}
By substituting these results back into equation (\ref{eq:vsunUN}) we get
\begin{align}
V_{SU(N)}=&2^{\frac{(N-1)(N-2)}{2}} N!\; \pi^\frac{N(N-1)}{2}\; 2^{\frac{N-1}{2}}
\pi^{(N-1)}\frac{1}{\sqrt{N!(N-1)!}}\;2^{-\frac{(N-2)(N-1)}{2}}\prod^{N-1}_{k=1}
\biggl(\frac{1}{k!}\biggr) \nonumber \\
=&2^{\frac{N-1}{2}}\pi^{\frac{(N-1)(N+2)}{2}}\sqrt{N}\prod^{N-1}_{k=1}
\biggl(\frac{1}{k!}\biggr)
\label{eq:eulermarinov}
\end{align}
which is just Marinov's initial formulation of the volume of $SU(N)$
\cite{Marinov, Marinov2}.  This is an important result, for it shows that the
overall generalized Euler angle parametrization of $SU(N)$ gives results
that are consistent with previously scrutinized work which used a completely different
methodology to derive the invariant volume of $SU(N)$.\footnote{Marinov 
calculated the invariant volumes for all compact simple
Lie groups by exploiting the spectral expansion of the Green's function for diffusion
on a group manifold \cite{Marinov}.}  

\section{N by N Density Matrix Parametrization}

We now turn out attention to the parametrization of
$N$ by $N$ density matrices.  We state that one may
parameterize any $N$ by $N$ density matrix as (see \cite{MByrd4All} for more details)
\begin{equation}
\label{rhodm}
\rho = U \rho_d U^\dagger,
\end{equation}
where $\rho_d$ is the diagonalized density matrix which corresponds 
to the eigenvalues of the ($N-1$)-sphere, $S^{N-1}$,
\begin{align}
\rho_d=&\left( \begin{matrix}
\sin^2(\theta_1)\cdots \sin^2(\theta_{N-1}) & 0 & \dots & 0 \\
0 & \cos^2(\theta_1)\cdots \sin^2(\theta_{N-1}) & \dots & 0 \\
\dots & \dots & \dots & \dots \\
0 & 0 & \dots & \cos^2(\theta_{N-1})
\end{matrix} \right)_{N\times N} \nonumber \\
=&\begin{pmatrix}
\Lambda_1, & 0 & \dots & 0 \\
0 & \Lambda_2 & \dots & 0 \\
\dots & \dots & \dots & \dots \\
0 & 0 & \dots & \Lambda_N 
\end{pmatrix}_{N \times N},
\end{align}
the range of the $\theta$ parameters is given by \cite{MByrd3Slater1, MByrd4All} 
\begin{equation}
 \cos^{-1}\biggl(\frac{1}{\sqrt{j+1}}\biggr) \leq \theta_j \leq \frac{\pi}{2},
\label{eq:rhodranges}
\end{equation}
and $U$ is from
equation (\ref{eq:suN}).  For completeness we should note that 
$U^\dagger$ is defined to be the conjugate
transpose of $U$, which, through equation (\ref{eq:suNconjeasF}), can be
seen to be equal to
\begin{align}
\label{eq:suNdagger}
U^\dagger=&\; e^{-i\lambda_{N^2-1} \alpha_{N^2-1}}e^{-i\lambda_{(N-1)^2-1}
\alpha_{N^2-2}} \cdots e^{-i\lambda_{3} \alpha_{N^2-(N-1)}} \nonumber \\
&\times \prod_{2 \leq m \leq N}\;\biggl(\prod_{m \geq k \geq 2}A(k,j(m))^\dagger \biggr),\nonumber \\
A(k,j(m))^\dagger =&\; e^{-i\lambda_{(k-1)^2+1} \alpha_{2(k-1)+j(m)}}e^{-i\lambda_{3} \alpha_{(2k-3)+j(m)}}, \nonumber \\
j(m) =&
\begin{cases}
0 \qquad &m=N,\\
\underset{0 \leq l \leq N-m-1}{\sum}2(m+l) \qquad &m \neq N. 
\end{cases}
\end{align}

Throughout the rest of the paper, $\rho_d$ will be parameterized by the following set 
of quantities \cite{MByrd2, MByrd3Slater1, MByrd4All, Tilma1}.
\begin{eqnarray}
\label{rhodform}
\rho_d &=& \frac{1}{N}\Bid_N + \sum_{2 \leq n \leq N} f(\theta_1, \theta_2, \ldots , \theta_{N-1})*\lambda_{n^2-1}, \nonumber\\
f(\theta_1, \theta_2, \ldots , \theta_{N-1})&=&\frac{1}{2}Tr[\rho_d \cdot \lambda_{n^2-1}].
\end{eqnarray}
For example, for the density matrix of two qubits $U$ is given by
equation (\ref{eq:su4eas}), $U^\dagger$ is
\begin{align}
\label{eq:su4easdagger}
U^\dagger =&\; e^{-i\lambda_{15} \alpha_{15}}e^{-i\lambda_8 \alpha_{14}}e^{-i\lambda_3 \alpha_{13}}e^{-i\lambda_2 \alpha_{12}}e^{-i\lambda_3 \alpha_{11}}e^{-i\lambda_5 \alpha_{10}}e^{-i\lambda_3 \alpha_9}e^{-i\lambda_2 \alpha_8}
\nonumber \\
&\times e^{-i\lambda_3 \alpha_7}e^{-i\lambda_{10} \alpha_6}e^{-i\lambda_3 \alpha_5}e^{-i\lambda_5 \alpha_4}e^{-i\lambda_3 \alpha_3}e^{-i\lambda_2 \alpha_2}e^{-i\lambda_3 \alpha_1},
\end{align}  
and $\rho_d$ is given by
\begin{eqnarray}
\rho_d &=& \left( \begin{matrix}
\sin^2(\theta_1)\sin^2(\theta_2)\sin^2(\theta_3) & 0 & 0 & 0\\
0 & \cos^2(\theta_1)\sin^2(\theta_2)\sin^2(\theta_3) & 0 & 0\\
0 & 0 & \cos^2(\theta_2)\sin^2(\theta_3) & 0 \\
0 & 0 & 0 & \cos^2(\theta_3) 
\end{matrix} \right) \nonumber \\
&=& \frac{1}{4}\Bid_4 + \sum_{2 \leq n \leq 4} f(\theta_1, \theta_2,
\theta_3)*\lambda_{n^2-1} \nonumber\\
&=& \frac{1}{4} \Bid_{4} + \frac{1}{2}(-1+2w^2)x^2y^2*\lambda_3 
+ \frac{1}{2\sqrt{3}}(-2+3x^2)y^2*\lambda_8 +
\frac{1}{2\sqrt{6}}(-3+4y^2)*\lambda_{15}
\end{eqnarray}
where 
\begin{gather}
w^2 = \sin^2(\theta_1), \qquad x^2 = \sin^2(\theta_2), \qquad y^2 = \sin^2(\theta_3), \nonumber \\ 
\frac{\pi}{4} \le \theta_1 \le \frac{\pi}{2}, \qquad
\cos^{-1}(\frac{1}{\sqrt{3}}) \le \theta_2 \le \frac{\pi}{2}, \qquad
\frac{\pi}{3} \le \theta_3 \le \frac{\pi}{2},
\end{gather}
and the one-quarter normalization of $\Bid_{4}$ 
keeps the trace of $\rho_d$ in this form still unity \cite{Tilma1}.
Therefore, using equations (\ref{eq:suN}), \eqref{rhodm},
(\ref{eq:suNdagger}), and \eqref{rhodform}, one can easily write down any $N$ by $N$ density matrix.

\section{Example: Haar Measures, Group Volumes and Density Matrices for 
Qubit/Qutrit, Three Qubit and Two Qutrit States}

Using the formalism we have now established, it is quite easy to write
down the Haar measure and group volume for $SU(8)$ and $SU(9)$, as well as the 
6 by 6, 8 by 8 and 9 by 9 dimensional density matrices, that
correspond to the qubit/qutrit, three qubit and two qutrit states.\footnote{
The Haar measure and group volume for $SU(6)$ has already been written down and calculated
in section \ref{sec:su6hm}.}

\subsubsection*{Case 1, Qubit/Qutrit states}
Here, a qubit interacts with a qutrit (a three-state system), 
thus yielding a 6-dimensional Hilbert
space which needs to be parameterized.  Using equations (\ref{eq:suN}),
\eqref{rhodm}, (\ref{eq:suNdagger}), and \eqref{rhodform} 
we arrive at the following formula for the density
matrix of a qubit/qutrit system:
\begin{align}
U=&\prod_{6 \geq m \geq 2}\;\biggl(\prod_{2 \leq k \leq m}
A(k,j(m))\biggr) \nonumber \\ 
&\times e^{i\lambda_{3} \alpha_{31}}e^{i\lambda_{8} \alpha_{32}}
e^{i\lambda_{15} \alpha_{33}}e^{i\lambda_{24} \alpha_{34}}e^{i\lambda_{35} \alpha_{35}},
\nonumber \\
A(k,j(m))=&e^{i\lambda_{3} \alpha_{(2k-3)+j(m)}}e^{i\lambda_{(k-1)^2+1} \alpha_{2(k-1)+j(m)}},
  \nonumber \\
U^\dagger=&\; e^{-i\lambda_{35} \alpha_{35}}
e^{-i\lambda_{24}\alpha_{34}} e^{-i\lambda_{15} \alpha_{33}} e^{-i\lambda_{8} \alpha_{32}}  e^{-i\lambda_{3} \alpha_{31}}\nonumber \\
&\times \prod_{2 \leq m \leq 6}\;\biggl(\prod_{m \geq k \geq 2}
A(k,j(m))^\dagger \biggr),\nonumber \\
A(k,j(m))^\dagger=&\; e^{-i\lambda_{(k-1)^2+1} \alpha_{2(k-1)+j(m)}}
e^{-i\lambda_{3} \alpha_{(2k-3)+j(m)}}, \nonumber \\
j(m) =&
\begin{cases}
0 \qquad &m=6,\\
\underset{0 \leq l \leq 6-m-1}{\sum}2(m+l) \qquad &m \neq 6,
\end{cases}\nonumber \\
\rho_d =& \frac{1}{6}\Bid_6 + \sum_{2 \leq n \leq 6} f(\theta_1, \theta_2, \ldots , \theta_{5})*\lambda_{n^2-1}.\label{eq:su6}
\end{align}
Explicitly:\begin{align}
\rho =&
\;e^{i\lambda_{3}\alpha_{1}}e^{i\lambda_{2}\alpha_{2}}
 e^{i\lambda_{3}\alpha_{3}}e^{i\lambda_{5}\alpha_{4}} 
 e^{i\lambda_{3}\alpha_{5}}e^{i\lambda_{10}\alpha_{6}}
 e^{i\lambda_{3}\alpha_{7}}e^{i\lambda_{17}\alpha_{8}} \nonumber \\
&\times e^{i\lambda_{3}\alpha_{9}}e^{i\lambda_{26}\alpha_{10}} 
 e^{i\lambda_{3}\alpha_{11}}e^{i\lambda_{2}\alpha_{12}} 
 e^{i\lambda_{3}\alpha_{13}}e^{i\lambda_{5}\alpha_{14}} 
 e^{i\lambda_{3}\alpha_{15}}e^{i\lambda_{10}\alpha_{16}} \nonumber \\
&\times e^{i\lambda_{3}\alpha_{17}}e^{i\lambda_{17}\alpha_{18}} 
 e^{i\lambda_{3}\alpha_{19}}e^{i\lambda_{2}\alpha_{20}}
 e^{i\lambda_{3}\alpha_{21}}e^{i\lambda_{5}\alpha_{22}} 
 e^{i\lambda_{3}\alpha_{23}}e^{i\lambda_{10}\alpha_{24}} \nonumber \\
&\times e^{i\lambda_{3}\alpha_{25}}e^{i\lambda_{2}\alpha_{26}} 
 e^{i\lambda_{3}\alpha_{27}}e^{i\lambda_{5}\alpha_{28}} 
 e^{i\lambda_{3}\alpha_{29}}e^{i\lambda_{2}\alpha_{30}}(\rho_d)
 e^{-i\lambda_{2}\alpha_{30}}e^{-i\lambda_{3}\alpha_{29}} \nonumber \\ 
&\times e^{-i\lambda_{5}\alpha_{28}}e^{-i\lambda_{3}\alpha_{27}}
 e^{-i\lambda_{2}\alpha_{26}}e^{-i\lambda_{3}\alpha_{25}} 
 e^{-i\lambda_{10}\alpha_{24}}e^{-i\lambda_{3}\alpha_{23}}
 e^{-i\lambda_{5}\alpha_{22}}e^{-i\lambda_{3}\alpha_{21}} \nonumber \\
&\times e^{-i\lambda_{2}\alpha_{20}}e^{-i\lambda_{3}\alpha_{19}}
 e^{-i\lambda_{17}\alpha_{18}}e^{-i\lambda_{3}\alpha_{17}}
 e^{-i\lambda_{10}\alpha_{16}}e^{-i\lambda_{3}\alpha_{15}}
 e^{-i\lambda_{5}\alpha_{14}}e^{-i\lambda_{3}\alpha_{13}}\nonumber \\
&\times e^{-i\lambda_{2}\alpha_{12}}e^{-i\lambda_{3}\alpha_{11}}
 e^{-i\lambda_{26}\alpha_{10}}e^{-i\lambda_{3}\alpha_{9}} 
 e^{-i\lambda_{17}\alpha_{8}}e^{-i\lambda_{3}\alpha_{7}} 
 e^{-i\lambda_{10}\alpha_{6}}e^{-i\lambda_{3}\alpha_{5}} \nonumber \\
&\times e^{-i\lambda_{5}\alpha_{4}}e^{-i\lambda_{3}\alpha_{3}}
 e^{-i\lambda_{2}\alpha_{2}}e^{-i\lambda_{3}\alpha_{1}},
\intertext{where}
\rho_d =&\frac{1}{6}\Bid_6 -\frac{ \cos (2\,{{\theta }_1})\,{\sin ({{\theta }_2})}^2\,{\sin ({{\theta }_3})}^2\,{\sin ({{\theta }_4})}^2\,{\sin ({{\theta }_5})}^2 }{2}*\lambda_3 \nonumber \\
&-\frac{ \left(1 + 3\,\cos (2\,{{\theta }_2}) \right) \,{\sin ({{\theta }_3})}^2\,{\sin ({{\theta }_4})}^2\,{\sin ({{\theta }_5})}^2 }{4\,{\sqrt{3}}}*\lambda_8 \nonumber \\
&-\frac{\left(1 + 2\,\cos (2\,{{\theta }_3}) \right) \,{\sin ({{\theta }_4})}^2\,{\sin ({{\theta }_5})}^2 }{2\,{\sqrt{6}}}*\lambda_{15} \nonumber \\
&-\frac{ \left(3 + 5\,\cos (2\,{{\theta }_4}) \right) \,{\sin ({{\theta }_5})}^2 }{4\,{\sqrt{10}}}*\lambda_{24} \nonumber \\
&-\frac{2 + 3\,\cos (2\,{{\theta }_5}) }{2\,{\sqrt{15}}}*\lambda_{35}.
\end{align}
From Appendix \ref{app:Haar} and equation (\ref{eq:rhodranges}) we know that the ranges for the quotient group $SU(6)/Z_6$ and the $\rho_d$ parameters are
\begin{align}
0 \leq \alpha_{2i-1} \leq \pi, \qquad 0 \leq \alpha_{2i} \leq \frac{\pi}{2}, \qquad (1 \leq i \leq 15) \nonumber \\
\cos^{-1}\biggl(\frac{1}{\sqrt{j+1}}\biggr) \leq \theta_j \leq \frac{\pi}{2}, \qquad (1 \leq j \leq 5).
\end{align}
And, for completeness, we again note that the Haar measure and group volume 
for a qubit/qutrit system has already been calculated in equation (\ref{eq:su6hmv}).

\subsubsection*{Case 2, Three qubit states}
Here, three qubits interact, thus yielding a 8-dimensional Hilbert
space which needs to be parameterized.  Using equations (\ref{eq:suN}),
\eqref{rhodm}, (\ref{eq:suNdagger}), and \eqref{rhodform} 
we arrive at the following formula for the density
matrix of the three qubit system:
\begin{align}
U=& \prod_{8 \geq m \geq 2}\;\biggl(\prod_{2 \leq k \leq m}
A(k,j(m))\biggr) \nonumber \\ 
&\times e^{i\lambda_{3} \alpha_{57}}e^{i\lambda_{8}
  \alpha_{58}} e^{i\lambda_{15} \alpha_{59}}e^{i\lambda_{24} \alpha_{60}} e^{i\lambda_{35} \alpha_{61}}e^{i\lambda_{48}
  \alpha_{62}} e^{i\lambda_{63} \alpha_{63}},
\nonumber \\
A(k,j(m))=&\; e^{i\lambda_{3}
  \alpha_{(2k-3)+j(m)}}e^{i\lambda_{(k-1)^2+1} \alpha_{2(k-1)+j(m)}},
  \nonumber \\
U^\dagger=&\; e^{-i\lambda_{63} \alpha_{63}}e^{-i\lambda_{48}
  \alpha_{62}}  e^{-i\lambda_{35} \alpha_{61}} 
  e^{-i\lambda_{24} \alpha_{60}}  e^{-i\lambda_{15} \alpha_{59}} e^{-i\lambda_{8}
  \alpha_{58}}  e^{-i\lambda_{3} \alpha_{57}} \nonumber \\
&\times \prod_{2 \leq m \leq 8}\;\biggl(\prod_{m \geq k \geq
  2}A(k,j(m))^\dagger \biggr),\nonumber \\
A(k,j(m))^\dagger=&\; e^{-i\lambda_{(k-1)^2+1} \alpha_{2(k-1)+j(m)}}e^{-i\lambda_{3}
  \alpha_{(2k-3)+j(m)}}, \nonumber \\
j(m) =&
\begin{cases}
0 \qquad &m=8,\\
\underset{0 \leq l \leq 8-m-1}{\sum}2(m+l) \qquad &m \neq 8,
\end{cases} \nonumber \\
\rho_d =& \frac{1}{8}\Bid_8 + \sum_{2 \leq n \leq 8} f(\theta_1, \theta_2, \ldots , \theta_{7})*\lambda_{n^2-1}.\label{eq:su8}
\end{align}
Explicitly: 
\begin{align}
\rho=&\; e^{i\lambda_{3} \alpha_{1}}e^{i\lambda_{2} \alpha_{2}}
e^{i\lambda_{3} \alpha_{3}}e^{i\lambda_{5} \alpha_{4}}
e^{i\lambda_{3} \alpha_{5}}e^{i\lambda_{10} \alpha_{6}} \nonumber \\
&\times e^{i\lambda_{3} \alpha_{7}}e^{i\lambda_{17} \alpha_{8}}
e^{i\lambda_{3} \alpha_{9}}e^{i\lambda_{26} \alpha_{10}}
e^{i\lambda_{3} \alpha_{11}}e^{i\lambda_{37} \alpha_{12}} \nonumber \\
&\times e^{i\lambda_{3} \alpha_{13}}e^{i\lambda_{50} \alpha_{14}}
e^{i\lambda_{3} \alpha_{15}}e^{i\lambda_{2} \alpha_{16}}
e^{i\lambda_{3} \alpha_{17}}e^{i\lambda_{5} \alpha_{18}} \nonumber \\
&\times e^{i\lambda_{3} \alpha_{19}}e^{i\lambda_{10} \alpha_{20}}
e^{i\lambda_{3} \alpha_{21}}e^{i\lambda_{17} \alpha_{22}}
e^{i\lambda_{3} \alpha_{23}}e^{i\lambda_{26} \alpha_{24}} \nonumber \\
&\times e^{i\lambda_{3} \alpha_{25}}e^{i\lambda_{37} \alpha_{26}}
e^{i\lambda_{3} \alpha_{27}}e^{i\lambda_{2} \alpha_{28}}
e^{i\lambda_{3} \alpha_{29}}e^{i\lambda_{5} \alpha_{30}} \nonumber \\
&\times e^{i\lambda_{3} \alpha_{31}}e^{i\lambda_{10} \alpha_{32}}
e^{i\lambda_{3} \alpha_{33}}e^{i\lambda_{17} \alpha_{34}}
e^{i\lambda_{3} \alpha_{35}}e^{i\lambda_{26} \alpha_{36}} \nonumber \\
&\times e^{i\lambda_{3} \alpha_{37}}e^{i\lambda_{2} \alpha_{38}}
e^{i\lambda_{3} \alpha_{39}}e^{i\lambda_{5} \alpha_{40}}
e^{i\lambda_{3} \alpha_{41}}e^{i\lambda_{10} \alpha_{42}} \nonumber \\
&\times e^{i\lambda_{3} \alpha_{43}}e^{i\lambda_{17} \alpha_{44}}
e^{i\lambda_{3} \alpha_{45}}e^{i\lambda_{2} \alpha_{46}}
e^{i\lambda_{3} \alpha_{47}}e^{i\lambda_{5} \alpha_{48}} \nonumber \\
&\times e^{i\lambda_{3} \alpha_{49}}e^{i\lambda_{10} \alpha_{50}}
e^{i\lambda_{3} \alpha_{51}}e^{i\lambda_{2} \alpha_{52}}
e^{i\lambda_{3} \alpha_{53}}e^{i\lambda_{5} \alpha_{54}} \nonumber \\
&\times e^{i\lambda_{3} \alpha_{55}}e^{i\lambda_{2} \alpha_{56}}(\rho_d)
e^{-i\lambda_{2} \alpha_{56}}e^{-i\lambda_{3} \alpha_{55}}
e^{-i\lambda_{5} \alpha_{54}}e^{-i\lambda_{3} \alpha_{53}} \\
&\times e^{-i\lambda_{2} \alpha_{52}}e^{-i\lambda_{3} \alpha_{51}}
e^{-i\lambda_{10} \alpha_{50}}e^{-i\lambda_{3} \alpha_{49}}
e^{-i\lambda_{5} \alpha_{48}}e^{-i\lambda_{3} \alpha_{47}} \nonumber \\
&\times e^{-i\lambda_{2} \alpha_{46}}e^{-i\lambda_{3} \alpha_{45}}
e^{-i\lambda_{17} \alpha_{44}}e^{-i\lambda_{3} \alpha_{43}}
e^{-i\lambda_{10} \alpha_{42}}e^{-i\lambda_{3} \alpha_{41}} \nonumber \\
&\times e^{-i\lambda_{5} \alpha_{40}}e^{-i\lambda_{3} \alpha_{39}}
e^{-i\lambda_{2} \alpha_{38}}e^{-i\lambda_{3} \alpha_{37}}
e^{-i\lambda_{26} \alpha_{36}}e^{-i\lambda_{3} \alpha_{35}} \nonumber \\
&\times e^{-i\lambda_{17} \alpha_{34}}e^{-i\lambda_{3} \alpha_{33}}
e^{-i\lambda_{10} \alpha_{32}}e^{-i\lambda_{3} \alpha_{31}}
e^{-i\lambda_{5} \alpha_{30}}e^{-i\lambda_{3} \alpha_{29}} \nonumber \\
&\times e^{-i\lambda_{2} \alpha_{28}}e^{-i\lambda_{3} \alpha_{27}}
e^{-i\lambda_{37} \alpha_{26}}e^{-i\lambda_{3} \alpha_{25}}
e^{-i\lambda_{26} \alpha_{24}}e^{-i\lambda_{3} \alpha_{23}} \nonumber \\
&\times e^{-i\lambda_{17} \alpha_{22}}e^{-i\lambda_{3} \alpha_{21}}
e^{-i\lambda_{10} \alpha_{20}}e^{-i\lambda_{3} \alpha_{19}}
e^{-i\lambda_{5} \alpha_{18}}e^{i\lambda_{3} \alpha_{17}} \nonumber \\
&\times e^{-i\lambda_{2} \alpha_{16}}e^{-i\lambda_{3} \alpha_{15}}
e^{-i\lambda_{50} \alpha_{14}}e^{-i\lambda_{3} \alpha_{13}}
e^{-i\lambda_{37} \alpha_{12}}e^{-i\lambda_{3} \alpha_{11}} \nonumber \\
&\times e^{-i\lambda_{26} \alpha_{10}}e^{-i\lambda_{3} \alpha_{9}}
e^{-i\lambda_{17} \alpha_{8}}e^{-i\lambda_{3} \alpha_{7}}
e^{-i\lambda_{10} \alpha_{6}}e^{-i\lambda_{3} \alpha_{5}} \nonumber \\
&\times e^{-i\lambda_{5} \alpha_{4}}e^{-i\lambda_{3} \alpha_{3}}
e^{-i\lambda_{2} \alpha_{2}}e^{-i\lambda_{3} \alpha_{1}}, \nonumber \\
\intertext{where}
\rho_d =&\frac{{{\Bid}_8}}{8}-\frac{\cos(2\,{{\theta}_1})\,{\sin({{\theta}_2})}^2\,{\sin({{\theta}_3})}^2\,
{\sin({{\theta}_4})}^2\,{\sin({{\theta}_5})}^2\,{\sin({{\theta}_6})}^2\,{\sin({{\theta}_7})}^2\,}{2}{{*\lambda}_3} \nonumber \\
&-\frac{\left(1+3\,\cos(2\,{{\theta}_2})\right)\,{\sin({{\theta}_3})}^2\,{\sin({{\theta}_4})}^2\,
{\sin({{\theta}_5})}^2\,{\sin({{\theta}_6})}^2\,{\sin({{\theta}_7})}^2\,}{4\,{\sqrt{3}}}{{*\lambda}_8} \nonumber \\
&-\frac{\left(1+2\,\cos(2\,{{\theta}_3})\right)\,{\sin({{\theta}_4})}^2\,{\sin({{\theta}_5})}^2\,
{\sin({{\theta}_6})}^2\,{\sin({{\theta}_7})}^2\,}{2\,{\sqrt{6}}}{{*\lambda}_{15}} \nonumber \\
&-\frac{\left(3+5\,\cos(2\,{{\theta}_4})\right)\,{\sin({{\theta}_5})}^2\,{\sin({{\theta}_6})}^2\,
{\sin({{\theta}_7})}^2\,}{4\,{\sqrt{10}}}{{*\lambda}_{24}} \nonumber \\
&-\frac{\left(2+3\,\cos(2\,{{\theta}_5})\right)\,{\sin({{\theta}_6})}^2\,{\sin({{\theta}_7})}^2\,}{2\,{\sqrt{15}}}{{*\lambda}_{35}} \nonumber \\
&-\frac{\left(5+7\,\cos(2\,{{\theta}_6})\right)\,{\sin({{\theta}_7})}^2\,}
{4\,{\sqrt{21}}}{{*\lambda}_{48}} \nonumber \\
&-\frac{\left(3+4\,\cos(2\,{{\theta}_7})\right)\,}{4\,{\sqrt{7}}}{{*\lambda}_{63}}.
\end{align}
From Appendix \ref{app:Haar} and equation (\ref{eq:rhodranges}) we know that the ranges for the quotient group $SU(8)/Z_8$ and the $\rho_d$ parameters are
\begin{align}
0 \leq \alpha_{2i-1} \leq \pi, \qquad 0 \leq \alpha_{2i} \leq \frac{\pi}{2}, \qquad (1 \leq i \leq 28) \nonumber \\
\cos^{-1}\biggl(\frac{1}{\sqrt{j+1}}\biggr) \leq \theta_j \leq \frac{\pi}{2}, \qquad (1 \leq j \leq 7),
\end{align}
combined with equations \eqref{dvKsun}
and (\ref{eq:dvSUN}), which yields the Haar measure for $SU(8)$
\begin{align}
dV_{SU(8)}=& \sin(2\alpha_2)\cos(\alpha_4)^3\sin(\alpha_4)\cos(\alpha_6)^5\sin(\alpha_6)
\cos(\alpha_8)^7\sin(\alpha_8)\cos(\alpha_{10})^9 \nonumber \\
&\times \sin(\alpha_{10})\cos(\alpha_{12})^{11}\sin(\alpha_{12})
\cos(\alpha_{14})\sin(\alpha_{14})^{13} \nonumber \\
&\times \sin(2\alpha_{16})\cos(\alpha_{18})^3\sin(\alpha_{18})\cos(\alpha_{20})^5\sin(\alpha_{20})
\cos(\alpha_{22})^7\sin(\alpha_{22})\cos(\alpha_{24})^9 \nonumber \\
&\times \sin(\alpha_{24})\cos(\alpha_{26})\sin(\alpha_{26})^{11}\nonumber \\
&\times \sin(2\alpha_{28})\cos(\alpha_{30})^3\sin(\alpha_{30})\cos(\alpha_{32})^5\sin(\alpha_{32})
\cos(\alpha_{34})^7\sin(\alpha_{34})\cos(\alpha_{36})\sin(\alpha_{36})^9 \nonumber \\
&\times \sin(2\alpha_{38})\cos(\alpha_{40})^3\sin(\alpha_{40})\cos(\alpha_{42})^5
\sin(\alpha_{42})\cos(\alpha_{44})\sin(\alpha_{44})^7 \nonumber \\
&\times \sin(2\alpha_{46})\cos(\alpha_{48})^3\sin(\alpha_{48})\cos(\alpha_{50})
\sin(\alpha_{50})^5 \nonumber \\
&\times \sin(2\alpha_{52})\cos(\alpha_{54})\sin(\alpha_{54})^3 \nonumber \\
&\times \sin(2\alpha_{56})d\alpha_{63}\ldots d\alpha_1,
\end{align}
one can calculate the group volume for the three qubit system;
\begin{align}
V_{SU(8)}=&2^{\frac{8-1}{2}}\pi^{\frac{(8-1)(8+2)}{2}}\sqrt{8}\prod^{8-1}_{k=1}
\biggl(\frac{1}{k!}\biggr) \nonumber \\
=&\frac{\pi^{35}}{391910400}.
\end{align}
which is what one gets with $N=8$ from equation (\ref{eq:eulermarinov}).

\subsubsection*{Case 3, Two Qutrits}
Here, two qutrits (two three-state systems) 
interact, thus yielding a 9-dimensional Hilbert
space which needs to be parameterized.  Using equations (\ref{eq:suN}),
\eqref{rhodm}, (\ref{eq:suNdagger}), and \eqref{rhodform} we arrive at the following formula 
for the density matrix for two qutrits:
\begin{align}
U=& \prod_{9 \geq m \geq 2}\;\biggl(\prod_{2 \leq k \leq m}
A(k,j(m))\biggr) \nonumber \\ 
&\times e^{i\lambda_{3} \alpha_{73}}e^{i\lambda_{8}
  \alpha_{74}} e^{i\lambda_{15} \alpha_{75}}e^{i\lambda_{24} \alpha_{76}} e^{i\lambda_{35} \alpha_{77}}e^{i\lambda_{48}
  \alpha_{78}} e^{i\lambda_{63} \alpha_{79}}e^{i\lambda_{80} \alpha_{80}},
\nonumber \\
A(k,j(m))=&\; e^{i\lambda_{3}
  \alpha_{(2k-3)+j(m)}}e^{i\lambda_{(k-1)^2+1} \alpha_{2(k-1)+j(m)}},
  \nonumber \\
U^\dagger=&\; e^{-i\lambda_{80} \alpha_{80}} e^{-i\lambda_{63} \alpha_{79}}e^{-i\lambda_{48}
  \alpha_{78}}  e^{-i\lambda_{35} \alpha_{77}}
  e^{-i\lambda_{24} \alpha_{76}}  e^{-i\lambda_{15} \alpha_{75}} e^{-i\lambda_{8}
  \alpha_{74}}  e^{-i\lambda_{3} \alpha_{73}} \nonumber \\
&\times \prod_{2 \leq m \leq 9}\;\biggl(\prod_{m \geq k \geq
  2}A(k,j(m))^\dagger \biggr),\nonumber \\
A(k,j(m))^\dagger=&\; e^{-i\lambda_{(k-1)^2+1} \alpha_{2(k-1)+j(m)}}e^{-i\lambda_{3}
  \alpha_{(2k-3)+j(m)}}, \nonumber \\
j(m) =&
\begin{cases}
0 \qquad &m=9,\\
\underset{0 \leq l \leq 9-m-1}{\sum}2(m+l) \qquad &m \neq 9,
\end{cases}\nonumber \\
\rho_d =& \frac{1}{9}\Bid_9 + \sum_{2 \leq n \leq 9} f(\theta_1, \theta_2,\ldots , \theta_{8})*\lambda_{n^2-1}. \label{eq:su9}
\end{align}
Explicitly: 
\begin{align}
\rho=&\; e^{i\lambda_{3} \alpha_{1}}e^{i\lambda_{2} \alpha_{2}}
e^{i\lambda_{3} \alpha_{3}}e^{i\lambda_{5} \alpha_{4}}
e^{i\lambda_{3} \alpha_{5}}e^{i\lambda_{10} \alpha_{6}} \nonumber \\
&\times e^{i\lambda_{3} \alpha_{7}}e^{i\lambda_{17} \alpha_{8}}
e^{i\lambda_{3} \alpha_{9}}e^{i\lambda_{26} \alpha_{10}}
e^{i\lambda_{3} \alpha_{11}}e^{i\lambda_{37} \alpha_{12}} \nonumber \\
&\times e^{i\lambda_{3} \alpha_{13}}e^{i\lambda_{50} \alpha_{14}}
e^{i\lambda_{3} \alpha_{15}}e^{i\lambda_{65} \alpha_{16}}
e^{i\lambda_{3} \alpha_{17}}e^{i\lambda_{2} \alpha_{18}} \nonumber \\
&\times e^{i\lambda_{3} \alpha_{19}}e^{i\lambda_{5} \alpha_{20}}
e^{i\lambda_{3} \alpha_{21}}e^{i\lambda_{10} \alpha_{22}}
e^{i\lambda_{3} \alpha_{23}}e^{i\lambda_{17} \alpha_{24}} \nonumber \\
&\times e^{i\lambda_{3} \alpha_{25}}e^{i\lambda_{26} \alpha_{26}}
e^{i\lambda_{3} \alpha_{27}}e^{i\lambda_{37} \alpha_{28}} 
e^{i\lambda_{3} \alpha_{29}}e^{i\lambda_{50} \alpha_{30}} \nonumber \\
&\times e^{i\lambda_{3} \alpha_{31}}e^{i\lambda_{2} \alpha_{32}}
e^{i\lambda_{3} \alpha_{33}}e^{i\lambda_{5} \alpha_{34}} 
e^{i\lambda_{3} \alpha_{35}}e^{i\lambda_{10} \alpha_{36}} \nonumber \\
&\times e^{i\lambda_{3} \alpha_{37}}e^{i\lambda_{17} \alpha_{38}}
e^{i\lambda_{3} \alpha_{39}}e^{i\lambda_{26} \alpha_{40}} 
e^{i\lambda_{3} \alpha_{41}}e^{i\lambda_{37} \alpha_{42}} \nonumber \\
&\times e^{i\lambda_{3} \alpha_{43}}e^{i\lambda_{2} \alpha_{44}}
e^{i\lambda_{3} \alpha_{45}}e^{i\lambda_{5} \alpha_{46}} 
e^{i\lambda_{3} \alpha_{47}}e^{i\lambda_{10} \alpha_{48}} \nonumber \\
&\times e^{i\lambda_{3} \alpha_{49}}e^{i\lambda_{17} \alpha_{50}}
e^{i\lambda_{3} \alpha_{51}}e^{i\lambda_{26} \alpha_{52}} 
e^{i\lambda_{3} \alpha_{53}}e^{i\lambda_{2} \alpha_{54}} \nonumber \\
&\times e^{i\lambda_{3} \alpha_{55}}e^{i\lambda_{5} \alpha_{56}}
e^{i\lambda_{3} \alpha_{57}}e^{i\lambda_{10} \alpha_{58}} 
e^{i\lambda_{3} \alpha_{59}}e^{i\lambda_{17} \alpha_{60}} \nonumber \\
&\times e^{i\lambda_{3} \alpha_{61}}e^{i\lambda_{2} \alpha_{62}}
e^{i\lambda_{3} \alpha_{63}}e^{i\lambda_{5} \alpha_{64}} 
e^{i\lambda_{3} \alpha_{65}}e^{i\lambda_{10} \alpha_{66}} \nonumber \\
&\times e^{i\lambda_{3} \alpha_{67}}e^{i\lambda_{2} \alpha_{68}}
e^{i\lambda_{3} \alpha_{69}}e^{i\lambda_{5} \alpha_{70}} 
e^{i\lambda_{3} \alpha_{71}}e^{i\lambda_{2} \alpha_{72}} \nonumber \\
&\times (\rho_d)
e^{-i\lambda_{2} \alpha_{72}}e^{-i\lambda_{3} \alpha_{71}}
e^{-i\lambda_{5} \alpha_{70}}e^{-i\lambda_{3} \alpha_{69}} \\
&\times e^{-i\lambda_{2} \alpha_{68}}e^{-i\lambda_{3} \alpha_{67}}
e^{-i\lambda_{10} \alpha_{66}}e^{-i\lambda_{3} \alpha_{65}}
e^{-i\lambda_{5} \alpha_{64}}e^{-i\lambda_{3} \alpha_{63}} \nonumber \\
&\times e^{-i\lambda_{2} \alpha_{62}}e^{-i\lambda_{3} \alpha_{61}}
e^{-i\lambda_{17} \alpha_{60}}e^{-i\lambda_{3} \alpha_{59}}
e^{-i\lambda_{10} \alpha_{58}}e^{-i\lambda_{3} \alpha_{57}} \nonumber \\
&\times e^{-i\lambda_{5} \alpha_{56}}e^{-i\lambda_{3} \alpha_{55}}
e^{-i\lambda_{2} \alpha_{54}}e^{-i\lambda_{3} \alpha_{53}}
e^{-i\lambda_{26} \alpha_{52}}e^{-i\lambda_{3} \alpha_{51}} \nonumber \\
&\times e^{-i\lambda_{17} \alpha_{50}}e^{-i\lambda_{3} \alpha_{49}}
e^{-i\lambda_{10} \alpha_{48}}e^{-i\lambda_{3} \alpha_{47}}
e^{-i\lambda_{5} \alpha_{46}}e^{-i\lambda_{3} \alpha_{45}} \nonumber \\
&\times e^{-i\lambda_{2} \alpha_{44}}e^{-i\lambda_{3} \alpha_{43}}
e^{-i\lambda_{37} \alpha_{42}}e^{-i\lambda_{3} \alpha_{41}}
e^{-i\lambda_{26} \alpha_{40}}e^{-i\lambda_{3} \alpha_{39}} \nonumber \\
&\times e^{-i\lambda_{17} \alpha_{38}}e^{-i\lambda_{3} \alpha_{37}}
e^{-i\lambda_{10} \alpha_{36}}e^{-i\lambda_{3} \alpha_{35}}
e^{-i\lambda_{5} \alpha_{34}}e^{i\lambda_{3} \alpha_{33}} \nonumber \\
&\times e^{-i\lambda_{2} \alpha_{32}}e^{-i\lambda_{3} \alpha_{31}}
e^{-i\lambda_{50} \alpha_{30}}e^{-i\lambda_{3} \alpha_{29}}
e^{-i\lambda_{37} \alpha_{28}}e^{-i\lambda_{3} \alpha_{27}} \nonumber \\
&\times e^{-i\lambda_{26} \alpha_{26}}e^{-i\lambda_{3} \alpha_{25}}
e^{-i\lambda_{17} \alpha_{22}}e^{-i\lambda_{3} \alpha_{21}}
e^{-i\lambda_{10} \alpha_{20}}e^{-i\lambda_{3} \alpha_{19}} \nonumber \\
&\times e^{-i\lambda_{5} \alpha_{18}}e^{-i\lambda_{3} \alpha_{17}}
e^{-i\lambda_{65} \alpha_{16}}e^{-i\lambda_{3} \alpha_{15}}
e^{-i\lambda_{50} \alpha_{14}}e^{-i\lambda_{3} \alpha_{13}} \nonumber \\
&\times e^{-i\lambda_{37} \alpha_{12}}e^{-i\lambda_{3} \alpha_{11}}
e^{-i\lambda_{26} \alpha_{10}}e^{-i\lambda_{3} \alpha_{9}}
e^{-i\lambda_{17} \alpha_{8}}e^{-i\lambda_{3} \alpha_{7}} \nonumber \\
&\times e^{-i\lambda_{10} \alpha_{6}}e^{-i\lambda_{3} \alpha_{5}}
e^{-i\lambda_{5} \alpha_{4}}e^{-i\lambda_{3} \alpha_{3}}
e^{-i\lambda_{2} \alpha_{2}}e^{-i\lambda_{3} \alpha_{1}}, \nonumber \\
\intertext{where}
\rho_d=&\frac{{{\Bid}_9}}{9}-\frac{\cos(2\,{{\theta}_1})\,{\sin({{\theta}_2})}^2\,
{\sin({{\theta}_3})}^2\,{\sin({{\theta}_4})}^2\,{\sin({{\theta}_5})}^2\,
{\sin({{\theta}_6})}^2\,{\sin({{\theta}_7})}^2\,{\sin({{\theta}_8})}^2\,}{2}{{*\lambda}_3}
\nonumber \\
&-\frac{\left(1+3\,\cos(2\,{{\theta}_2})\right)\,{\sin({{\theta}_3})}^2\,
{\sin({{\theta}_4})}^2\,{\sin({{\theta}_5})}^2\,{\sin({{\theta}_6})}^2\,
{\sin({{\theta}_7})}^2\,{\sin({{\theta}_8})}^2\,}{4\,{\sqrt{3}}}{{*\lambda}_8}
\nonumber \\
&-\frac{\left(1+2\,\cos(2\,{{\theta}_3})\right)\,{\sin({{\theta}_4})}^2\,
{\sin({{\theta}_5})}^2\,{\sin({{\theta}_6})}^2\,{\sin({{\theta}_7})}^2\,
{\sin({{\theta}_8})}^2\,}{2\,{\sqrt{6}}}{{*\lambda}_{15}}\nonumber \\
&-\frac{\left(3+5\,\cos(2\,{{\theta}_4})\right)\,{\sin({{\theta}_5})}^2\,
{\sin({{\theta}_6})}^2\,{\sin({{\theta}_7})}^2\,{\sin({{\theta}_8})}^2\,}{4\,{\sqrt{10}}}{{*\lambda}_{24}}
\nonumber \\
&-\frac{\left(2+3\,\cos(2\,{{\theta}_5})\right)\,
{\sin({{\theta}_6})}^2\,{\sin({{\theta}_7})}^2\,{\sin({{\theta}_8})}^2\,}{2\,{\sqrt{15}}}{{*\lambda}_{35}}
\nonumber \\
&-\frac{\left(5+7\,\cos(2\,{{\theta}_6})\right)\,
{\sin({{\theta}_7})}^2\,{\sin({{\theta}_8})}^2\,}{4\,{\sqrt{21}}}{{*\lambda}_{48}}
\nonumber \\
&-\frac{\left(3+4\,\cos(2\,{{\theta}_7})\right)\,{\sin({{\theta}_8})}^2\,}{4\,{\sqrt{7}}}{{*\lambda}_{63}}
\nonumber \\
&-\frac{\left(\frac{7}{12}-\frac{3\,\cos(2\,{{\theta}_8})}{4}\right)\,
}{2}{{*\lambda}_{80}}.
\end{align}
From Appendix \ref{app:Haar} and equation (\ref{eq:rhodranges}) we know that the ranges for the quotient group $SU(9)/Z_9$ and the $\rho_d$ parameters are
\begin{align}
0 \leq \alpha_{2i-1} \leq \pi, \qquad 0 \leq \alpha_{2i} \leq \frac{\pi}{2}, \qquad (1 \leq i \leq 36) \nonumber \\
\cos^{-1}\biggl(\frac{1}{\sqrt{j+1}}\biggr) \leq \theta_j \leq \frac{\pi}{2}, \qquad (1 \leq j \leq 8),
\end{align}
combined with equations \eqref{dvKsun}
and (\ref{eq:dvSUN}), which yields the Haar measure for $SU(9)$
\begin{align}
dV_{SU(9)}=& \sin(2\alpha_2)\cos(\alpha_4)^3\sin(\alpha_4)\cos(\alpha_6)^5\sin(\alpha_6)
\cos(\alpha_8)^7\sin(\alpha_8)\cos(\alpha_{10})^9 \nonumber \\
&\times \sin(\alpha_{10})\cos(\alpha_{12})^{11}\sin(\alpha_{12})
\cos(\alpha_{14})^{13}\sin(\alpha_{14})\cos(\alpha_{16})\sin(\alpha_{16})^{15} \nonumber \\
&\times\sin(2\alpha_{18})\cos(\alpha_{20})^3\sin(\alpha_{20})\cos(\alpha_{22})^5\sin(\alpha_{22})
\cos(\alpha_{24})^7\sin(\alpha_{24})\cos(\alpha_{26})^9 \nonumber \\
&\times \sin(\alpha_{26})\cos(\alpha_{28})^{11}\sin(\alpha_{28})
\cos(\alpha_{30})\sin(\alpha_{30})^{13} \nonumber \\
&\times \sin(2\alpha_{32})\cos(\alpha_{34})^3\sin(\alpha_{34})\cos(\alpha_{36})^5\sin(\alpha_{36})
\cos(\alpha_{38})^7\sin(\alpha_{38})\cos(\alpha_{40})^9 \nonumber \\
&\times \sin(\alpha_{40})\cos(\alpha_{42})\sin(\alpha_{42})^{11}\nonumber \\
&\times \sin(2\alpha_{44})\cos(\alpha_{46})^3\sin(\alpha_{46})\cos(\alpha_{48})^5\sin(\alpha_{48})
\cos(\alpha_{50})^7\sin(\alpha_{50})\cos(\alpha_{52})\sin(\alpha_{52})^9 \nonumber \\
&\times \sin(2\alpha_{54})\cos(\alpha_{56})^3\sin(\alpha_{56})\cos(\alpha_{58})^5
\sin(\alpha_{58})\cos(\alpha_{60})\sin(\alpha_{60})^7 \nonumber \\
&\times \sin(2\alpha_{62})\cos(\alpha_{64})^3\sin(\alpha_{64})\cos(\alpha_{66})
\sin(\alpha_{66})^5 \nonumber \\
&\times \sin(2\alpha_{68})\cos(\alpha_{70})\sin(\alpha_{70})^3 \nonumber \\
&\times \sin(2\alpha_{72})d\alpha_{80}\ldots d\alpha_1,
\end{align}
one can calculate the group volume for the two qutrit system;
\begin{align}
V_{SU(9)}=&2^{\frac{9-1}{2}}\pi^{\frac{(9-1)(9+2)}{2}}\sqrt{9}\prod^{9-1}_{k=1} 
\biggl(\frac{1}{k!}\biggr) \nonumber \\
=&\frac{\pi^{44}}{105345515520000}.
\end{align}
which is what one gets with $N=9$ from equation (\ref{eq:eulermarinov}).

\section{Conclusions/Comments}

The aim of this paper has been to show an explicit Euler angle 
parametrization for $SU(N)$ as well as the Hilbert space for all $N$ by $N$ density matrices.\footnote{This is but one
parametrization that can be realized for a unitary group.  For example Dr. Vilenkin and 
Dr. Klimyk in \textit{Representation of Lie Groups and 
Special Functions} \cite{VK} have an extensive treatment of the unitary and orthogonal groups using a similar treatment
as was used here.}  
This parametrization also yields a general form for the Haar
measure for $SU(N)$ as well as confirms Marinov's initial 
group volume formula for $SU(N)$.
It should be noted that such a parametrization could 
be very useful in numerous areas of study, most notably 
numerical calculations concerning entanglement and other
quantum information parameters.\footnote{This has already occurred for the $SU(4)$ case, see
for example \texttt{quant-ph$\backslash$0203088} and \texttt{quant-ph$\backslash$0207181}.}
This parametrization may also allow for an in-depth 
analysis of the convex sets, sub-sets, and overall set boundaries of
separable and entangled qubit, qutrit, and $N$-trit systems without having to make any
initial restrictions as to the type of 
parametrization and density matrix in question.  

\section{Acknowledgments}

We would like to thank Dr.\ M.\ Byrd for introducing the
parametrization problem, as well as Kim Hojeong for helping develop the
initial parametrization for $SU(N)$.  
\pagebreak

\appendix

\section{Diagram Formalism for $SU(N)$ Parametrization}
\label{app:pictureformalism}
The invocation of ``removal of the redundancies'' 
in order to achieve the necessary $N^2-1$ number of parameters in our $SU(N)$ 
parametrization seems
to be an arbitrary addition to an otherwise rigorous mathematical
discussion.  This assumed arbitrariness is due, in part, to an
underlying construction that is the basis for the formulation of the
Euler angle parametrization that we have given.  The goal of this
appendix is to describe that construction and to alleviate any
remaining sense of arbitrariness in the development of the parametrization.\footnote{Using diagrams
to develop a representation of a group is a standard tool in developing group parametrizations.  An 
excellent example of this is Dr. Vilenkin and Dr. Klimyk \textit{Representation of Lie Groups and 
Special Functions} section 10.5 \cite{VK}.} 

To begin, let's take a two-component column matrix \textbf{w} given by
\begin{equation}
\text{\textbf{w}}=\begin{pmatrix}
a\\
b \end{pmatrix}.
\end{equation}
We wish to find the minimal number of operations that rotates, or
``mixes'', the two components.  Since \textbf{w} is a 2\;x\;1 column matrix, 
our operations
must be 2\;x\;2 square matrices in order to generate a 2\;x\;1 column matrix
as output.  These 2\;x\;2 square matrices are going to be elements of
$SU(2)$.  Therefore we should look at the well known Euler angle
parametrization of $SU(2)$ to find these operations.  

Recall that $SU(2)$ is the covering group of $SO(3)$, the group of
rotations of three dimensional Euclidean space, with the following representation:
\begin{quote}
Every rotation $R\in SO(3)$ can be parameterized by an axis of rotation
$\hat{n}$ and the angle $\theta$ of rotation about the
axis: $R=(\hat{n},\theta)$.  The axis requires two angles, $(\alpha,
\beta)$, for its specification so three parameters are needed to
specify a general rotation: $SO(3)$ is a three parameter
group \cite{Sattinger}.
\end{quote}
Now, any arbitrary rotation $R$ in
Euclidean space can be accomplished in three steps, known as Euler
rotations, which are characterized by three angles, known as
Euler angles.  When one wants an arbitrary rotation in (x,y,z) space, 
one then does the following:
\begin{enumerate}
\item{Rotate through an angle $\alpha$ about the z-axis.}
\item{Rotate through an angle $\beta$ about the new 
y-axis.\footnote{In Classical Mechanics, this rotation is conventionally 
done about the new x-axis \cite{CorbenStehle}.}}
\item{Rotate through an angle $\gamma$ about the new z-axis.}
\end{enumerate}
The rotation matrix $R$ describing these operations is thus given by 
\begin{equation}
\label{bfR}
R(\alpha, \beta, \gamma) =
R_{z_{new}}(\gamma)R_{y_{new}}(\beta)R_z(\alpha),
\end{equation}
where
\begin{equation}
R_z(\alpha)=
\begin{pmatrix}
\cos[\alpha] & \sin[\alpha] & 0 \\
-\sin[\alpha] & \cos[\alpha] & 0 \\
0 & 0 & 1 
\end{pmatrix},
\quad
R_{y_{new}}(\beta)=
\begin{pmatrix}
1 & 0 & 0 \\
0 & \cos[\beta] & \sin[\beta] \\
0 & -\sin[\beta] & \cos[\beta] 
\end{pmatrix},
\quad
R_{z_{new}}(\gamma)=
\begin{pmatrix}
\cos[\alpha] & \sin[\alpha] & 0 \\
-\sin[\alpha] & \cos[\alpha] & 0 \\
0 & 0 & 1 
\end{pmatrix}.
\end{equation}

Unfortunately at this point we have to make a distinction between body-fixed (the
new axes) and space-fixed (the original axes) coordinates.  For a
general rotation, we would like to represent $R(\alpha, \beta, \gamma)$ in
terms of space-fixed axes \cite{Sattinger, CorbenStehle} and not body-fixed 
axes.\footnote{This distinction comes from the physical origins of the three-dimensional 
rotation group: the rotations of a rigid body about a fixed point 
constitute the group which we now know to be $SO(3)$.  The terminology 
is from the physical distinction between the coordinate frame 
``attached'' to the rigid body, and the coordinate frame of the 
``surrounding'' system \cite{CorbenStehle}.}  This is can be achieved in two steps.
First one makes the
following definitions \cite{WignerNotes}
\begin{equation}
R_{y_{new}}(\beta)=R_z(\alpha)R_y(\beta)(R_z(\alpha))^{-1}
\quad 
\text{ and }
\quad
R_{z_{new}}(\gamma)=R_{y_{new}}(\beta)R_z(\gamma)(R_{y_{new}}(\beta))^{-1}.
\end{equation}
Then, substituting these definitions back into the body-fixed version of $R(\alpha, \beta, \gamma)$ given in equation \eqref{bfR} we get 
\begin{align}
R(\alpha, \beta,
\gamma)=&\;R_{z_{new}}(\gamma)R_{y_{new}}(\beta)R_z(\alpha) \nonumber \\
=&\;R_{y_{new}}(\beta)R_z(\gamma)(R_{y_{new}}(\beta))^{-1}R_{y_{new}}(\beta)R_z(\alpha) \nonumber \\
=&\;R_{y_{new}}(\beta)R_z(\gamma)R_z(\alpha) \nonumber \\
=&\;R_z(\alpha)R_y(\beta)(R_z(\alpha))^{-1}R_z(\gamma)R_z(\alpha)
\nonumber \\
=&\;R_z(\alpha)R_y(\beta)R_z(\gamma) 
\end{align}
where we define the y-axis rotation as
\begin{equation}
R_y(\beta)=\begin{pmatrix}
\cos[\beta] & 0 & \sin[\beta] \\
0 & 1 & 0 \\
-\sin[\beta] & 0 & \cos[\beta]
\end{pmatrix}
\end{equation}
and the last step in the derivation was accomplished by 
noticing that rotations about
the same axis commute.

Thus, one generally does a rotation about the
z axis, then about the y axis, then another rotation about the z axis,
in order to describe the most general rotation of a rigid body.  But,
if one is rather interested in the ``mixing'' of the components of a
Euclidean three-vector describing a point in the rigid body (from the point of view
of the space-fixed axes), then a rotation about the 
z axis is equivalent to a ``mixing''
between the first and second components of the three vector, the
rotation about the y axis is equivalent to a ``mixing'' of the first and
third components, and the second z axis rotation, can be seen to be
yet another ``mixing'' of the first and second components.\footnote{We will see a
similar discussion when we look at elements of $SU(3)$ acting on a three
component column matrix.}  If we look at 
the second z axis rotation as imparting an overall phase, then the
first two rotations become the key ``mixing'' rotations while the third rotation 
imparts an overall phase to the components of the three-vector.\footnote{The 
fact that we are having the second z-axis rotation impart nothing more than 
an overall phase is due in part to an understanding that this 
geometric approach to the parametrization is not to be taken as 
as a completely rigorous mathematical derivation, but rather as an aid in visualizing the 
important group elements of the $SU(N)$ Euler angle parametrization (which when combined 
with the correspondingly appropriate exponentiated 
Cartan subalgebra components at the end of the 
representation, generates both the correct number of parameters for the group,
as well as correctly imparts the needed overall phase onto ``mixed''
column matrix components).  
For example, in the $SU(2)$ parametrization case, the third rotation is
considered part of the Cartan subalgebra of the parametrization, and can therefore be
ignored until later in the parametrization's development.}

In conclusion then, we need the relationship between the rotation matrix
$R(\alpha, \beta, \gamma)$ and it's $SU(2)$ counterpart in order to accomplish the desired
``mixing'' of \textbf{w}.  This relationship comes from the
following mappings between the $SO(3)$ matrix elements and the $SU(2)$
matrix elements (see for example \cite{Greiner,Artin,Sattinger} for
more details)
\begin{eqnarray}
\pm \begin{pmatrix}
\cos[\frac{\alpha}{2}] & i\sin[\frac{\alpha}{2}] \\
i\sin[\frac{\alpha}{2}] & \cos[\frac{\alpha}{2}]
\end{pmatrix}
&\Longrightarrow
\begin{pmatrix}
1 & 0 & 0 \\
0 & \cos[\alpha] & - \sin[\alpha] \\
0 & \sin[\alpha] & \cos[\alpha]
\end{pmatrix} \nonumber \\
\pm \begin{pmatrix}
\cos[\frac{\beta}{2}] & -\sin[\frac{\beta}{2}] \\
-\sin[\frac{\beta}{2}] & \cos[\frac{\beta}{2}]
\end{pmatrix}
&\Longrightarrow
\begin{pmatrix}
\cos[\beta] & 0 & \sin[\beta] \\
0 & 1 & 0 \\
-\sin[\beta] & 0 & \cos[\beta]
\end{pmatrix} \nonumber \\
\pm \begin{pmatrix}
e^{i\frac{\gamma}{2}} & 0 \\
0 & e^{-i\frac{\gamma}{2}}
\end{pmatrix}
&\Longrightarrow
\begin{pmatrix}
\cos[\gamma] & -\sin[\gamma] & 0 \\
\sin[\gamma] & \cos[\gamma] & 0 \\
0 & 0 & 1
\end{pmatrix}
\label{su2so3}
\end{eqnarray}
From these mappings we see that for a $U \in SU(2)$, an equivalent
rotation matrix to $R(\alpha, \beta, \gamma)$ can be written
as \footnote{The equivalence is that $\theta = \frac{\alpha}{2}$, $\phi
  = \frac{\beta}{2}$, and $\psi = \frac{\gamma}{2}$.}
\begin{align}
U =&\; e^{i\sigma_3 \theta}e^{i\sigma_2 \phi}e^{i\sigma_3 \psi}
\nonumber \\
=& \begin{pmatrix}
e^{i(\theta+\psi)}\cos[\phi] & e^{i(\theta-\psi)}\sin[\phi] \\
-e^{i(-\theta+\psi)}\sin[\phi] & e^{-i(\theta+\psi)}\cos[\phi]
\end{pmatrix}.
\label{eq:su2mixer}
\end{align}
If we act upon \textbf{w} with equation (\ref{eq:su2mixer}) we
generate the rotated column matrix \textbf{W}
\begin{equation}
\label{wrotated}
\text{\textbf{W}} = \begin{pmatrix}
a*e^{i(\theta+\psi)}\cos[\phi]+b*e^{i(\theta-\psi)}\sin[\phi] \\
b*e^{-i(\theta+\psi)}\cos[\phi]-a*e^{i(-\theta+\psi)}\sin[\phi]
\end{pmatrix}.
\end{equation}
Therefore, following the previous discussion concerning $SO(3)$
rotations, a 
simplified rotated, or ``mixed'' form of \textbf{w} can be seen to be given by
\begin{align}
\text{\textbf{W}} =&\; e^{i\sigma_3 \theta}e^{i\sigma_2 \phi} \text{\textbf{w}} \nonumber \\
=& \begin{pmatrix}
a*e^{i\theta}\cos[\phi]+b*e^{i\theta}\sin[\phi] \\
b*e^{-i\theta}\cos[\phi]-a*e^{-i\theta}\sin[\phi]
\end{pmatrix}.
\label{eq:simplewrotated}
\end{align}
This reduced $SU(2)$ group operation generating \textbf{W} can be pictorially
represented by the following diagram
\begin{equation}
\label{su2groupdiagram}
\begin{bmatrix}
a\\
b
\end{bmatrix} \equiv e^{i\sigma_3 \theta}e^{i\sigma_2 \phi}.
\end{equation}
With this pictorially representation of the group action which ``mixes''
column matrix components, we can now graphically express and decompose 
$N$-component column matrix ``mixings.''

We can start using this graphical method by looking at a three 
component column matrix.  
Here what we want is the minimal
number of operations that will effectively ``mix'' the three component
column matrix \textbf{v}
\begin{equation}
\text{\textbf{v}}=\begin{pmatrix}
a\\
b\\
c \end{pmatrix}.
\end{equation}
Comparing with the previous work, we notice that 
since \textbf{w} is a 3\;x\;1 column matrix, our operations
must be 3\;x\;3 square matrices in order to generate the necessary 3\;x\;1 
column matrix
as output.  These 3\;x\;3 square matrices are going to be elements of
$SU(3)$.  Therefore we should look at the Euler angle parametrization of
$SU(3)$, initially given in
\cite{MByrd1, MByrdp1} and later in \cite{Gibbons}, to find these operations.

Before we address the full $SU(3)$ group, the necessary rotations of
\textbf{v} can be expressed by extending our diagram representation of
the $SU(2)$ rotations in the following way
\begin{equation}
\label{eq:su3diagram} 
\begin{bmatrix}
a\\
b\\
c
\end{bmatrix} =\;
\begin{bmatrix}
a\\
b
\end{bmatrix} \text{ then }
\biggl(\begin{bmatrix}
a\\
b
\end{bmatrix} \text{ and }
\begin{bmatrix}
a\\
\\
c
\end{bmatrix}\biggr),
\end{equation}
where
\begin{equation}
\biggl(\begin{bmatrix}
a\\
b
\end{bmatrix} \text{ and }
\begin{bmatrix}
a\\
\\
c
\end{bmatrix}\biggr) =\;
\begin{bmatrix}
a\\
\\
c
\end{bmatrix} \text{ then }
\begin{bmatrix}
a\\
b
\end{bmatrix}.
\end{equation}
Since we wish to rotate, or
``mix'', components a and b and components b and c, we can see that this can be
done in stages; first by mixing components a and b then by mixing a and b
and a and c together.  Therefore, what is needed to accomplish this scheme is the group element of 
$SU(3)$ that will rotate components a and c, without rotating b
(although an overall phase on b is acceptable).  This
action is represented by the following $SU(3)$ element
\begin{equation}
\label{su3rotationelement}
e^{i\lambda_5 \zeta}=\begin{pmatrix}
\cos[\zeta] & 0 & \sin[\zeta] \\
0 & 1 & 0 \\
-\sin[\zeta] & 0 & \cos[\zeta] \end{pmatrix}.
\end{equation}
With equation \eqref{su3rotationelement} and following the methodology
from the $SU(2)$ work, 
we can associate the previously diagrammed $SU(3)$
rotation element with the following group element from $SU(3)$
\begin{equation}
\begin{bmatrix}
a\\
\\
c
\end{bmatrix} \equiv \; e^{i\lambda_3 \eta}e^{i\lambda_5 \zeta}.
\footnote{The reason we have a $\lambda_3$ and not a $\lambda_8$ 
in the first group element is beyond the scope of this Appendix, but is 
addressed in the following Appendices.}
\end{equation}
Thus the $SU(3)$ rotations, diagrammed in equation (\ref{eq:su3diagram}),
are seen to be equivalent to the following group action \footnote{In this 
representation we are using the general extension of the fact that the ``product'' 
of two rotations \textit{$R_1$} and \textit{$R_2$}, denoted \textit{$R_1R_2$} 
is the transformation resulting from acting \textit{first} 
with $R_2$, \textit{then} with $R_1$ \cite{Biedenharn}.}
\begin{align}
\begin{bmatrix}
a\\
b\\
c
\end{bmatrix} =& \; \begin{bmatrix}
a\\
b
\end{bmatrix} \cdot
\begin{bmatrix}
a\\
\\
c
\end{bmatrix} \cdot
\begin{bmatrix}
a\\
b
\end{bmatrix} \nonumber \\
\equiv & \; e^{i\lambda_3 \theta_2}e^{i\lambda_2 \phi_2}e^{i\lambda_3
\eta}e^{i\lambda_5 \zeta}e^{i\lambda_3 \theta}e^{i\lambda_2\phi}.
\label{eq:su3groupdiagram}
\end{align}

The extension of this work to ``mixing'' four component, and thus, n-component
column matrices is straightforward.  For a four component column matrix \textbf{t}
\begin{equation}
\text{\textbf{t}}=\begin{pmatrix}
a\\
b\\
c\\
d \end{pmatrix}
\end{equation}
we schematically represent the necessary rotations as 
\begin{equation}
\label{eq:su4diagram}
\begin{bmatrix}
a\\
b\\
c\\
d
\end{bmatrix} =\;
\begin{bmatrix}
a\\
b
\end{bmatrix} \text{ then }
\biggl(\begin{bmatrix}
a\\
b
\end{bmatrix} \text{ and }
\begin{bmatrix}
a\\
\\
c
\end{bmatrix}\biggr) \text{ then }
\biggl(\begin{bmatrix}
a\\
b
\end{bmatrix} \text{ and }
\begin{bmatrix}
a\\
\\
c
\end{bmatrix} \text{ and }
\begin{bmatrix}
a\\
\\
\\
d
\end{bmatrix}\biggr)
\end{equation}
where
\begin{equation}
\biggl(\begin{bmatrix}
a\\
b
\end{bmatrix} \text{ and }
\begin{bmatrix}
a\\
\\
c
\end{bmatrix} \text{ and }
\begin{bmatrix}
a\\
\\
\\
d
\end{bmatrix}\biggr) =\;
\begin{bmatrix}
a\\
\\
\\
d
\end{bmatrix} \text{ then }
\begin{bmatrix}
a\\
\\
c
\end{bmatrix} \text{ then }
\begin{bmatrix}
a\\
b
\end{bmatrix}.
\end{equation}
Following the pattern of the previous work, the rotation between 
components a and d is achieved by acting
upon \textbf{t} with a 4\;x\;4 matrix which is going to be a $SU(4)$ group element
\begin{equation}
\label{su4rotationelement}
e^{i\lambda_{10} \chi}=\begin{pmatrix}
\cos[\chi] & 0 & 0 & \sin[\chi] \\
0 & 1 & 0 & 0 \\
0 & 0 & 1 & 0 \\
-\sin[\chi] & 0 & 0 & \cos[\chi] \end{pmatrix}.
\end{equation}
Mathematically then, the above ``mixing'' scheme can be seen to be
given by the following group action \footnote{See previous footnote.}
\begin{align}
\begin{bmatrix}
a\\
b\\
c\\
d
\end{bmatrix} =& \; 
\begin{bmatrix}
a\\
b
\end{bmatrix} \cdot
\begin{bmatrix}
a\\
\\
c
\end{bmatrix} \cdot
\begin{bmatrix}
a\\
\\
\\
d
\end{bmatrix} \cdot
\begin{bmatrix}
a\\
b
\end{bmatrix} \cdot
\begin{bmatrix}
a\\
\\
c
\end{bmatrix} \cdot
\begin{bmatrix}
a\\
b
\end{bmatrix} \nonumber \\
\equiv & \; e^{i\lambda_3 \theta_3}e^{i\lambda_2 \phi_3}
e^{i\lambda_3\eta_2}e^{i\lambda_5 \zeta_2}e^{i\lambda_3 \epsilon}
e^{i\lambda_{10}\chi}e^{i\lambda_3 \theta_2}e^{i\lambda_2 \phi_2}
e^{i\lambda_3\eta}e^{i\lambda_5 \zeta}e^{i\lambda_3 \theta}e^{i\lambda_2\phi}.
\label{eq:su4groupdiagram}
\end{align}

In general then, our graphical representation of ``mixing'' column matrix
components comes from both Biedenharn's work on angular 
momentum \cite{Biedenharn} and 
from a conceptual extension of 
the Clebsch-Gordon coefficient construction for spin $\frac{1}{2}$ 
particles \cite{WignerNotes}.  Recall that for spin $\frac{1}{2}$ particles, 
the only part of the $SU(2)$ rotation 
operator which mixes different quantum number $m$ values in the Clebsch-Gordon 
coefficients is the middle rotation; the first and third add nothing but 
an overall phase.  As we have 
seen, that middle rotation corresponds to the $\sigma_2$ Pauli Matrix for $SU(2)$ 
and to the $\lambda_{(N-1)^2+1}$ matrix for $SU(N)$, $N>2$. It is these Lie algebra 
components that when exponentiated, ``mix'' the column matrix 
components in question.  One then only needs to ``attach'' (matrix multiply) 
the appropriate exponentiation of the Cartan 
subalgebra ($\sigma_3$ for $SU(2)$, $\lambda_3$ and $\lambda_8$ 
for $SU(3)$, and $\lambda_3, \lambda_8$ and $\lambda_{15}$ for $SU(4)$) at then 
end of the group action (given in equation \eqref{su2groupdiagram} for $SU(2)$, 
equation (\ref{eq:su3groupdiagram}) for $SU(3)$, and 
equation (\ref{eq:su4groupdiagram}) for $SU(4)$) to 
achieve the necessary overall phase for the column matrix 
components.  The generalization of this methodology to 
higher $SU(N)$ groups should now be quite apparent.  
\pagebreak

\section{Invariant Volume Element Normalization Calculations}
\label{app:Haar}
Before integrating $dV_{SU(N)}$ we need some group theory.  
We begin with a digression concerning the center of a
group \cite{Artin, Scott} .  If $S$ is a subset of a group $G$, then the centralizer,
$C_{G}(S)$ of $S$ in $G$ is defined by
\begin{equation}
C(S) \equiv C_{G}(S) = \{x\in G \mid \text{ if } s\in S \text{ then } xs = sx\}.
\end{equation}
For example, if $S=\{y\}$, $C(y)$ will be used instead of $C(\{y\})$.  Next,
the centralizer of $G$ in $G$ is called the center of $G$ and is denoted by
$Z(G)$ or $Z$.  
\begin{equation}
\begin{split}
Z(G) \equiv Z&= \{z\in G \mid zx=xz  \text{ for all } x\in G\}\\
&=C_{G}(G).
\end{split}
\end{equation}
Another way of writing
this is
\begin{equation}
\begin{split}
Z(G) &= \cap \{C(x) \mid x\in G\}\\
& = \{z \mid \text{if } x\in G \text{ then } z\in C(x)\}.
\end{split}
\end{equation}
In other words, the center is the set of all elements $z$ that commutes
with all other elements in the group.  Finally, the commutator $[x,y]$ of two elements $x$ and $y$ of a group $G$ is
given by the equation
\begin{equation}
[x,y]=x^{-1}y^{-1}xy.
\end{equation}

Now what we want to find is the number of elements in the center of
$SU(N)$ for $N=2, 3, 4$ and so on.  Begin by defining the
following
\begin{equation}
Z_{n}=\text{ cyclic group of order n }\cong \mathbb{Z}_{n} \cong Z(SU(N)).
\end{equation}
Therefore, the set of all matrices which comprise the center of $SU(N)$, $Z(SU(N))$, is congruent to $Z_{N}$ since
we know that if $G$ is a finite linear group over a field \textit{F}, then the set of matrices of
the form $\Sigma c_{g}g$, where $g \in G$ and $c_g \in F$, forms an algebra (in
fact, a ring) \cite{Scott, Sattinger}.
For example, for $SU(2)$ we would have
\begin{equation}
\begin{split}
Z_{2} =& \{x\in SU(2) \mid [x,y]\in Z_1 \text{ for all } y\in SU(2)\},\\
[x,y] =&\omega \Bid_2,\\
Z_1 =&\{\Bid_2\}.
\end{split}
\end{equation}
This would be the set of all 2 by 2 matrix elements such that the commutator
relationship would yield the identity matrix multiplied by some
non-zero coefficient.  In general this can be
written as
\begin{equation}
\begin{split}
Z_{N} =& \{x\in SU(N) \mid [x,y]\in Z_1 \text{ for all } y\in SU(N)\},\\
Z_1 =&\{\Bid_N\}.
\end{split}
\end{equation}
This is similar to the result from \cite{Artin}, that shows that the 
center of the general linear
group of real matrices, $GL_{N}(\Re)$, is the group of scalar
matrices, that is, those of the from 
$\omega \mathbb{I}$, where $\mathbb{I}$ is the identity
element of the group and $\omega$ is some multiplicative constant.  For $SU(N)$, 
$\omega \mathbb{I}$ is an $N^{th}$ root of unity.   

To begin our actual search for the normalization constant for our invariant group 
element, we first again look at the group $SU(2)$.  For this group, every element can be
written as
\begin{equation}
\begin{pmatrix}
a & b\\
-\bar{b} & \bar{a}
\end{pmatrix}
\end{equation}
where $|a|^2 + |b|^2=1$.  Again, following \cite{Artin} we can make the
following parametrization
\begin{equation}
\begin{split}
a &= y_1 - i y_2,\\
b &= y_3 - i y_4,\\
1 &= y_{1}^2 + y_{2}^2 + y_{3}^2 + y_{4}^2.
\end{split}
\end{equation}
The elements $(1,0,0,0)$ and $(-1,0,0,0)$ are anti-podal points, 
or polar points if one pictures the
group as a three-dimensional unit sphere in a 4-dimensional space
parameterized by y, and
thus comprise the elements for the center group of $SU(2)$ (i.e. $\pm
\Bid_2$).  Therefore the center of $SU(2)$, $Z_2$, is comprised
of two elements; $\pm \Bid_2$.

Now, in our parametrization, the general $SU(2)$ elements are given by
\begin{equation}
\label{su2vol}
\begin{split}
D(\mu, \nu, \xi) &= e^{i\lambda_{3}\mu}e^{i\lambda_{2}\nu}e^{i\lambda_{3}\xi},\\  
dV_{SU(2)} &= \sin(2\nu)d\mu d\nu d\xi,
\end{split}
\end{equation}
with corresponding ranges
\begin{gather}
0 \le \mu,\xi \le \pi,\\
0 \le \nu \le \frac{\pi}{2}.
\end{gather}
Integrating over the volume element $dV_{SU(2)}$ with the above
ranges yields the volume of the group $SU(2)/Z_{2}$.  In other
words, the $SU(2)$ group modulo its center $Z_{2}$.  In
order to get the full volume of the $SU(2)$ group, all ones need to do
is multiply the volume of $SU(2)/Z_{2}$ by the number of identified
center elements; in this case two.

This process can be extended to the $SU(3)$ and $SU(4)$ parametrizations.  
For $SU(3)$ \cite{MByrd1, MByrdp1, MByrd2, MByrd3Slater1}
(here recast as a component of the $SU(4)$ parametrization derived in
\cite{Tilma1})
\begin{equation}
SU(3) = e^{i\lambda_3\alpha_7}e^{i\lambda_2\alpha_8}e^{i\lambda_3\alpha_9}
e^{i\lambda_{5}\alpha_{10}}D(\alpha_{11}, \alpha_{12}, \alpha_{13})
e^{i\lambda_{8}\alpha_{14}}.
\end{equation}
Now, we get an initial factor of two from the $D(\alpha_{11}, \alpha_{12}, \alpha_{13})$
component.  We shall now proves 
that we get another factor of two from the $e^{i\lambda_3\alpha_9}
e^{i\lambda_{5}\alpha_{10}}$ component as well.   

From the 
commutation relations of the elements of 
the Lie algebra of $SU(3)$ (see \cite{MByrd1} for details) 
we see that $\{\lambda_3, \lambda_4, \lambda_5, \lambda_8\}$ form a closed subalgebra
$SU(2)\times U(1)$.\footnote{Georgi \cite{Georgi} has stated that $\lambda_2, \lambda_5, \text{ and } \lambda_7$ 
generate an $SU(2)$ subalgebra of $SU(3)$.  This fact can be seen in the commutator
relationships between these three $\lambda$ matrices contained in \cite{Tilma1} or in \cite{MByrd1}.} 
\begin{align}
[\lambda_3, \lambda_4]=&i\lambda_5, \nonumber \\
[\lambda_3, \lambda_5]=&-i\lambda_4, \nonumber \\
[\lambda_3, \lambda_8]=&0, \nonumber \\
[\lambda_4, \lambda_5]=&i(\lambda_3+\sqrt{3}\lambda_8), \nonumber \\
[\lambda_4, \lambda_8]=&-i\sqrt{3}\lambda_5, \nonumber \\
[\lambda_5, \lambda_8]=&i\sqrt{3}\lambda_4.
\end{align}
Observation of the four $\lambda$ matrices with respect 
to the Pauli spin matrices of $SU(2)$ shows that $\lambda_4$ is the $SU(3)$ analogue of 
$\sigma_1$, $\lambda_5$ is the $SU(3)$ analogue of $\sigma_2$ and $\lambda_8$ is 
the $SU(3)$ analogue of $\sigma_3$
\begin{equation}
\begin{aligned}
\sigma_1&=\begin{pmatrix}
0 & 1 \\
1 & 0
\end{pmatrix}
\quad \Longrightarrow \quad
\lambda_4=\begin{pmatrix}
0 & 0 & 1 \\
0 & 0 & 0 \\
1 & 0 & 0 
\end{pmatrix}, \\
\sigma_2&=\begin{pmatrix}
0 & -i \\
i & 0
\end{pmatrix}
\quad \Longrightarrow \quad
\lambda_5=\begin{pmatrix}
0 & 0 & -i \\
0 & 0 & 0 \\
i & 0 & 0 
\end{pmatrix}, \\
\sigma_3&=\begin{pmatrix}
1 & 0 \\
0 & -1
\end{pmatrix}
\quad \Longrightarrow \quad
\lambda_3=\begin{pmatrix}
1 &  0 & 0 \\
0 & -1 & 0 \\
0 &  0 & 0
\end{pmatrix}
\text{ and }
\lambda_8=\frac{1}{\sqrt{3}}\begin{pmatrix}
1 & 0 & 0 \\
0 & 1 & 0 \\
0 & 0 & -2 
\end{pmatrix}.
\end{aligned}
\end{equation}
Thus one may use either $\{\lambda_3, \lambda_5 \}$ 
or $\{\lambda_3, \lambda_5, \lambda_8 \}$ to generate an $SU(2)$ subgroup of 
$SU(3)$.  The volume of this $SU(2)$ subgroup of $SU(3)$ must be equal to the volume of the
general $SU(2)$ group; $2\pi^2$.
If we demand that any element of the $SU(2)$ 
subgroup of $SU(3)$ have similar ranges as its $SU(2)$ 
analogue\footnote{This requires a normalization
factor of $\frac{1}{\sqrt{3}}$ on the maximal range of $\lambda_8$ that is explained
by the removal of the $Z_3$ elements of $SU(3)$.}, then a multiplicative factor of 2 is required for the 
$e^{i\lambda_3\alpha_9}e^{i\lambda_{5}\alpha_{10}}$ component.\footnote{When calculating this volume 
element, it is important to remember that the closed subalgebra being used is 
$SU(2)\times U(1)$ and therefore the integrated kernel, be it derived either from 
$e^{i\lambda_3 \alpha}e^{i\lambda_5 \beta}e^{i\lambda_3 \gamma}$ or 
$e^{i\lambda_3 \alpha}e^{i\lambda_5 \beta}e^{i\lambda_8 \gamma}$, will require contributions from both
the $SU(2)$ and $U(1)$ elements.}

Finally, $SU(3)$ has a $Z_{3}$ whose elements have the generic form 
\begin{equation}
\begin{pmatrix}
\eta_1 & 0 & 0\\
0 & \eta_2 & 0\\
0 & 0 & \eta_1^{-1}\eta_2^{-1}
\end{pmatrix},
\end{equation}
where
\begin{equation}
\eta_1^3 = \eta_2^3 = 1.
\end{equation}
Solving for $\eta_1$ and $\eta_2$ yields the following elements for $Z_{3}$
\begin{equation}
\begin{pmatrix}
                     1 & 0 & 0 \\
                     0 & 1 & 0 \\
                     0 & 0 & 1 
\end{pmatrix}, \quad
\begin{array}{cc}
-\left( \begin{array}{ccc} 
                     (-1)^\frac{1}{3} & 0 & 0 \\
                     0 & (-1)^\frac{1}{3} & 0 \\
                     0 & 0 & (-1)^\frac{1}{3} \end{array} \right),
                     \quad
\left( \begin{array}{ccc} 
                     (-1)^\frac{2}{3} & 0 & 0 \\
                     0 & (-1)^\frac{2}{3} & 0 \\
                     0 & 0 & (-1)^\frac{2}{3}\end{array} \right)
\end{array}
\end{equation}
which are the three cube roots of unity.  Combining these $SU(3)$ center elements, a
total of three, with the 2 factors of 2 from the previous discussion, 
yields a total multiplication factor of 12.  The volume
of $SU(3)$ is then
\begin{eqnarray}
V_{SU(3)} &=& 2*2*3*V(SU(3)/Z_3) \nonumber\\
&=& \sqrt{3}\,\pi^5
\end{eqnarray}
using the ranges given above for the general $SU(2)$ elements, combined with 
$0 \le \alpha_{14} \le \frac{\pi}{\sqrt{3}}$.  Explicitly:
\begin{gather}
0 \le \alpha_7, \alpha_9, \alpha_{11}, \alpha_{13} \le \pi, \nonumber \\
0 \le \alpha_8, \alpha_{10}, \alpha_{12} \le \frac{\pi}{2}, \nonumber \\
0 \le \alpha_{14} \le \frac{\pi}{\sqrt{3}}.
\end{gather}
These are modifications of \cite{MByrd1, MByrdp1, MByrd2, MByrd3Slater1, MByrdp2} and take into account the
updated Marinov group volume values \cite{Marinov2}.

For $SU(4)$ the process is similar to that used for $SU(3)$, but now with two $SU(2)$ subgroups
to worry about.  For a $U \in SU(4)$, the derivation of which can be found in \cite{Tilma1}, 
we see that, 
\begin{equation}
U = e^{i\lambda_3 \alpha_1}e^{i\lambda_2 \alpha_2}e^{i\lambda_3 \alpha_3}e^{i\lambda_5 \alpha_4}e^{i\lambda_3 \alpha_5}e^{i\lambda_{10} \alpha_6}[SU(3)]e^{i\lambda_{15} \alpha_{15}}.
\end{equation}
Here, the two $SU(2)$ subalgebras in $SU(4)$ that we are concerned with 
are $\{\lambda_3, \lambda_4, \lambda_5, \lambda_8, \lambda_{15}\}$ and
$\{\lambda_3, \lambda_9, \lambda_{10}, \lambda_8, \lambda_{15}\}$.
Both of these $SU(2)\times U(1)\times U(1)$ subalgebras are 
represented in the parametrization of $SU(4)$ as $SU(2)$ subgroup elements, 
$e^{i\lambda_3 \alpha_3}e^{i\lambda_5 \alpha_4}$ and 
$e^{i\lambda_3 \alpha_5}e^{i\lambda_{10} \alpha_6}$.  We can see that 
$\lambda_{10}$ 
is the $SU(4)$ analogue of $\sigma_2$\footnote{We have already discussed 
$\lambda_5$ in the previous section on $SU(3)$.} and $\lambda_{15}$ is the 
$SU(4)$ analogue to $\sigma_3$\footnote{It is the $SU(4)$ Cartan subalgebra element.}.
The demand that all $SU(2)$ subgroups of $SU(4)$ must have a volume
equal to $2\pi^2$ is equivalent to having 
the parameters of the associated elements of the $SU(2)$ subgroup 
run through similar ranges as their $SU(2)$ analogues.\footnote{This requires a normalization
factor of $\frac{1}{\sqrt{6}}$ on the maximal range of $\lambda_{15}$ that is explained
by the removal of the $Z_4$ elements of $SU(4)$.}  
As with $SU(3)$, this restriction yields an overall multiplicative factor of 4 from 
these two elements.\footnote{When calculating these volume 
elements, it is important to remember that the closed subalgebra being used is 
$SU(2)\times U(1)\times U(1)$ and therefore, as in the $SU(3)$ case, 
the integrated kernels will 
require contributions from appropriate Cartan subalgebra elements.  For example, the 
$e^{i\lambda_3\alpha_3}e^{i\lambda_{5}\alpha_4}$ component is an 
$SU(2)$ sub-element of the parametrization of $SU(4)$,
but in creating its corresponding $SU(2)$ subgroup volume kernel (see the $SU(3)$ discussion), 
one must remember that it is a $SU(2)\subset SU(3) \subset SU(4)$ 
and therefore the kernel only requires 
contributions from the $\lambda_3$ and $\lambda_8$ components. On the other hand, the
$e^{i\lambda_3\alpha_5}e^{i\lambda_{10}\alpha_6}$ element corresponds to a 
$SU(2)\subset SU(4)$ and
therefore, the volume kernel will require contributions from all
three Cartan subalgebra elements of $SU(4)$.}  Recalling that the $SU(3)$ element yields a 
multiplicative factor of 12, all that remains is 
to determine the multiplicative factor equivalent to the 
identification of the $SU(4)$ center, $Z_{4}$.  

The elements of the center of $SU(4)$ are similar in form to the ones from $SU(3)$;
\begin{equation}
\begin{pmatrix}
\eta_1 & 0 & 0 & 0\\
0 & \eta_2 & 0 & 0\\
0 & 0 & \eta_3 & 0\\
0 & 0 & 0 & \eta_1^{-1}\eta_2^{-1}\eta_3^{-1}
\end{pmatrix},
\end{equation}
where
\begin{equation}
\eta_1^4 = \eta_2^4 = \eta_3^4 = 1.
\end{equation}
Solving yields the 4 roots of unity: 
$
\pm \Bid_4 \text{ and } \pm i \Bid_4,
$
where $\Bid_4$ is the 4\;x\;4 identity matrix.
So we can see that $Z_{4}$ gives another factor of 4, which, when 
combined with the factor of 4 from the two $SU(2)$ subgroups,
and the factor of 12 from the $SU(3)$ elements, gives a total 
multiplicative factor of 192.  

Thus, when we integrate the $SU(4)$ differential volume element with the
ranges given previously for the general $SU(2)$ and $SU(3)$ elements combined
with the appropriate range for the $\lambda_{15}$ component (all combined
below)
\begin{gather}
0 \le \alpha_1, \alpha_3, \alpha_5, 
\alpha_7, \alpha_9, \alpha_{11}, \alpha_{13} \le \pi, \nonumber \\
0 \le \alpha_2, \alpha_4, \alpha_6,
\alpha_8, \alpha_{10}, \alpha_{12} \le \frac{\pi}{2}, \nonumber \\
0 \le \alpha_{14} \le \frac{\pi}{\sqrt{3}}, \nonumber \\
0 \le \alpha_{15} \le \frac{\pi}{\sqrt{6}},
\end{gather}
we get
\begin{eqnarray}
V_{SU(4)} &=& 2*2*2*2*3*4*V(SU(4)/Z_4)\nonumber\\
&=&\frac{\sqrt{2}\,\pi^9}{3}.
\end{eqnarray}  
This calculated volume for $SU(4)$, the derivation of which can be found in \cite{Tilma1} 
agrees with that from Marinov \cite{Marinov2}.

From this work, it is plain to see that in general, the ranges
for the $\lambda_3$, $\lambda_2$ analogues (recall equation
(\ref{eq:suNmat})), and the remaining Cartan subalgebra components 
of the parametrization will take the following general form
\begin{gather}
0 \le \alpha(\lambda_3) \le \pi, \nonumber \\
0 \le \alpha(\lambda_{(k-1)^2+1}) \le \frac{\pi}{2}, \nonumber \\
0 \le \alpha(\text{ Cartan subalgebra elements }) \le \pi
\sqrt{\frac{2}{k(k-1)}},
\label{eq:genranges}
\end{gather}
for $2\le k \le N$.\footnote{An observant reader will notice that 
we have counted $\lambda_3$ twice in the above ranges.  The fact of
the matter is that we wanted to distinguish between the 
first and second $\lambda_3$ elements in the fundamental $SU(2)$ 
parametrization.  The first is given as the $\alpha(\lambda_3)$
component while the second is regarded as the $SU(2)$ Cartan subalgebra
component.  The reason for this distinction will be 
made clearer in Appendix \ref{app:paramranges}.} 
Also, it is apparent that the elements of
the center of $SU(N)$ will have the form
\begin{equation}
\begin{pmatrix}
\eta_1 & 0 & 0 & 0 & \dots & 0\\
0 & \eta_2 & 0 & 0 & \dots & 0\\
0 & 0 & \eta_3 & 0 & \dots & 0\\
\dots & \dots & \dots & \dots & \dots & \dots \\
0 & 0 & 0 & 0 & \dots &
\eta_1^{-1}\eta_2^{-1}\eta_3^{-1} \ldots \eta_{N-1}^{-1}
\end{pmatrix}
\end{equation}
where
\begin{equation}
\eta_1^N = \eta_2^N = \eta_3^N = \cdots = \eta_{N-1}^N = 1,
\end{equation}
thus yielding $N$ elements in $Z(SU(N))$ and a corresponding
multiplicative constant of $N$ in the calculation of the group volume.
Observation of the previous calculations of the invariant volume
element for $SU(2)$, $SU(3)$, and $SU(4)$ also indicates that the Euler
angle parametrization for $SU(N)$ yields $(N-2)$ $SU(2)$ subgroups that
require multiplication by two in order to satisfy the $2\pi^2$
general $SU(2)$ volume requirement.  
Therefore, if one defines the ranges given in
equation (\ref{eq:genranges}) as $V'$, we see that the invariant volume
element for $SU(N)$ can be written as
\begin{equation}
V_{SU(N)}= \Omega_{N}*\idotsint\limits_{V^\prime}
dV_{SU(N)}
\end{equation}
where
\begin{equation}
\Omega_{N}=2^{N-2}*N*\Omega_{N-1}
\end{equation}
since the differential volume element $dV_{SU(N)}$,
given in \eqref{dvsun}, shows a  
reliance on the differential volume element of $SU(N-1)$ and therefore
on $\Omega_{N-1}$. 
\pagebreak

\section{Modified Parameter Ranges for Group Covering}
\label{app:paramranges}
In order to be complete, we list the modifications to the ranges given
in Appendix \ref{app:Haar} that affect a covering of $SU(2)$, $SU(3)$, $SU(4)$, and $SU(N)$ 
in general
without jeopardizing the calculated group volumes.

To begin, in our parametrization, the general $SU(2)$ elements are given by
\begin{equation}
\begin{split}
D(\mu, \nu, \xi) &= e^{i\lambda_{3}\mu}e^{i\lambda_{2}\nu}e^{i\lambda_{3}\xi},\\  
dV_{SU(2)} &= \sin(2\nu)d\mu d\nu d\xi,
\end{split}
\end{equation}
with the corresponding ranges for the volume of $SU(2)/Z_2$ given as
\begin{gather}
0 \le \mu,\xi \le \pi,\nonumber \\
0 \le \nu \le \frac{\pi}{2}.
\end{gather}
In order to generate a covering of $SU(2)$, the $\xi$ parameter must be
modified to take into account the uniqueness of the two central group
elements, $\pm \Bid_2$, under spinor transformations.  This modification
is straightforward enough; $\xi$'s range is multiplied by the number
of central group elements in $SU(2)$.  The new ranges are thus 
\begin{gather}
0 \le \mu \le \pi,\nonumber \\
0 \le \nu \le \frac{\pi}{2},\nonumber \\
0 \le \xi \le 2\pi.
\end{gather}
These ranges yield both a covering of $SU(2)$, as well as the correct
group volume for $SU(2)$.\footnote{One may interchange $\mu$ and $\xi$'s ranges without altering either 
the volume calculation, or the final orientation of a two-vector under operation by $D$.  This interchange
is beneficial when looking at Euler parametrizations beyond $SU(2)$.}  

One can also see this by looking at the
values of the finite group elements under both sets of ranges.  To do
this, we first partition $D(\mu, \nu, \xi)$ as 
\begin{equation}
D((\mu, \nu), \xi)=(e^{i\lambda_{3}\mu}e^{i\lambda_{2}\nu})e^{i\lambda_{3}\xi},
\end{equation}
where
\begin{eqnarray}
e^{i\lambda_{3}\mu} &=& \begin{pmatrix}
e^{i\mu} & 0 \\
0 & e^{-i\mu} \\
\end{pmatrix}, \nonumber \\
e^{i\lambda_{2}\nu} &=& \begin{pmatrix}
cos(\nu) & sin(\nu) \\
-sin(\nu) & cos(\nu) \\
\end{pmatrix}, \\
e^{i\lambda_{3}\xi} &=& \begin{pmatrix}
e^{i\xi} & 0 \\
0 & e^{-i\xi} \\
\end{pmatrix}. \nonumber
\end{eqnarray}
Then looking at $0 \le \xi \le 2\pi$ first we see the following pattern
emerge
\begin{align}
\xi=&\;0 \Longrightarrow  \begin{pmatrix}
1 & 0 \\
0 & 1 \\
\end{pmatrix} \Longrightarrow \Bid_{2}, \nonumber \\
\xi=&\;\pi \Longrightarrow  \begin{pmatrix}
-1 & 0 \\
0 & -1 \\
\end{pmatrix} \Longrightarrow -\Bid_{2}, \nonumber \\
\xi=&\;2\pi \Longrightarrow  \begin{pmatrix}
1 & 0 \\
0 & 1 \\
\end{pmatrix} \Longrightarrow \Bid_{2} \Longleftarrow \xi=0. \quad
\text{\textbf{Repeat}}
\end{align}
Next, we look at $0 \le \mu \le \pi$ and $0 \le \nu \le \frac{\pi}{2}$
\begin{align}
\mu=0 \;,&\; \nu=0 \Longrightarrow \Bid_{2}, \nonumber \\
\mu=\pi \;,&\; \nu=0 \Longrightarrow -\Bid_{2}, \nonumber \\
\mu=0 \;,&\; \nu=\frac{\pi}{2} \Longrightarrow \begin{pmatrix}
0 & 1 \\
-1 & 0 \\
\end{pmatrix}, \nonumber \\
\mu=\pi \;,&\; \nu=\frac{\pi}{2} \Longrightarrow \begin{pmatrix}
0 & -1 \\
1 & 0 \\
\end{pmatrix}.
\label{eq:su232fe}
\end{align}
Combining these two results yields the following 12 matrices for
$D((\mu, \nu), \xi)$
\begin{eqnarray}
&D((0,0),0)=\left( \begin{array}{cc} 
                     1 & 0 \\
                     0 & 1 \end{array} \right), \quad
&D((\pi,0),0)=-\left(\begin{array}{cc} 
                     1 & 0 \\
                     0 & 1 \end{array} \right), \nonumber \\
&D((0,0),\pi)=-\left( \begin{array}{cc} 
                     1 & 0 \\
                     0 & 1 \end{array} \right), \quad
&D((\pi,0),\pi)=\left(\begin{array}{cc} 
                     1 & 0 \\
                     0 & 1 \end{array} \right), \nonumber \\
&D((0,0),2\pi)=\left( \begin{array}{cc} 
                     1 & 0 \\
                     0 & 1 \end{array} \right), \quad
&D((\pi,0),2\pi)=-\left(\begin{array}{cc} 
                     1 & 0 \\
                     0 & 1 \end{array} \right), \nonumber \\
&D((0,\frac{\pi}{2}),0)=\left( \begin{array}{cc} 
                     0 & 1 \\
                     -1 & 0 \end{array} \right), \quad
&D((\pi,\frac{\pi}{2}),0)=\left( \begin{array}{cc} 
                     0 & -1 \\
                     1 & 0 \end{array} \right), \nonumber \\
&D((0,\frac{\pi}{2}),\pi)=\left( \begin{array}{cc} 
                     0 & -1 \\
                     1 & 0 \end{array} \right), \quad
&D((\pi,\frac{\pi}{2}),\pi)=\left( \begin{array}{cc} 
                     0 & 1 \\
                     -1 & 0 \end{array} \right), \nonumber \\
&D((0,\frac{\pi}{2}),2\pi)=\left( \begin{array}{cc} 
                     0 & 1 \\
                     -1 & 0 \end{array} \right), \quad
&D((\pi,\frac{\pi}{2}),2\pi)=\left( \begin{array}{cc} 
                     0 & -1 \\
                     1 & 0 \end{array} \right).
\end{eqnarray}
We can see that in the above 12 matrices we have repeated four
fundamental forms, each three times
\begin{eqnarray}
&\Bid_{2} \Longrightarrow 3, \qquad \begin{pmatrix}
0 & 1 \\
-1 & 0 \\
\end{pmatrix}  \Longrightarrow 3, \nonumber \\
&-\Bid_{2} \Longrightarrow 3, \qquad \begin{pmatrix}
0 & -1 \\
1 & 0 \\
\end{pmatrix} \Longrightarrow 3,
\end{eqnarray}
when we have run $\xi$ from $0$ to $2\pi$.  But, if we remove the
repeating $\Bid_{2}$ term in the original $\xi$ calculations above -
in essence forcing $\xi$ to range from $0$ to $\pi$ instead of $2\pi$
- we see that we then get the same four fundamental forms, but now
each only repeated twice
\begin{eqnarray}
&\Bid_{2} \Longrightarrow 2, \qquad \begin{pmatrix}
0 & 1 \\
-1 & 0 \\
\end{pmatrix}\Longrightarrow 2, \nonumber \\
&-\Bid_{2} \Longrightarrow 2, \qquad \begin{pmatrix}
0 & -1 \\
1 & 0 \\
\end{pmatrix} \Longrightarrow 2.
\end{eqnarray}
The greatest common divisor of the above list is, obviously, 2, which
not only corresponds to the amount that the range of the $\xi$ parameter was
divided by, but also to the multiplicative factor of 2
that is required in the calculation of the invariant volume element
when using the quotient group ranges.  This may seem trivial, but let
us now look at $SU(3)$.

For $SU(3)$, here given as a component of the $SU(4)$ parametrization,
we know we have one $SU(2)$ subgroup component (from Appendix \ref{app:Haar}) as well
as the $SU(2)$ contribution $D(\mu, \nu, \xi)$ here rewritten in terms
of the $SU(3)$ parameters $\alpha_{11}, \text{ } \alpha_{12}, \text{ and
  } \alpha_{13}$:
\begin{equation}
SU(3) = e^{i\lambda_3\alpha_7}e^{i\lambda_2\alpha_8}e^{i\lambda_3\alpha_9}
e^{i\lambda_{5}\alpha_{10}}D(\alpha_{11}, \alpha_{12}, \alpha_{13})
e^{i\lambda_{8}\alpha_{14}}.
\end{equation}
Therefore the ranges of $\alpha_9$ and $\alpha_{13}$ should be modified
just as $\xi$'s was done in the previous discussion for
$SU(2)$.\footnote{Recall the previous footnote on the interchange of the
first and third component ranges in a $SU(2)$ parametrization.} 
Remembering the discussion in
Appendix \ref{app:Haar} concerning the central group of $SU(3)$, we can deduce that
$\alpha_{14}$'s ranges should be multiplied by a factor of 3.   This
yields the following, corrected, ranges for $SU(3)$\footnote{Earlier
representations of these ranges for $SU(3)$, for example in
\cite{MByrd1, MByrdp1, MByrd2, MByrd3Slater1, MByrdp2}, were
incorrect in that they failed to take into account the updated $SU(N)$ volume formula in \cite{Marinov2}.}
\begin{gather}
0 \le \alpha_7,\alpha_{11} \le \pi,\nonumber \\
0 \le \alpha_8,\alpha_{10},\alpha_{12} \le \frac{\pi}{2},\nonumber \\
0 \le \alpha_9,\alpha_{13} \le 2\pi,\nonumber \\
0 \le \alpha_{14} \le \sqrt{3}\pi.
\end{gather}
These ranges yield both a covering of $SU(3)$, as well as the correct
group volume for $SU(3)$.

For a $U \in SU(4)$, we have two $SU(2)$ subgroup components
\begin{equation}
U = e^{i\lambda_3 \alpha_1}e^{i\lambda_2 \alpha_2}e^{i\lambda_3 \alpha_3}e^{i\lambda_5 \alpha_4}e^{i\lambda_3 \alpha_5}e^{i\lambda_{10} \alpha_6}[SU(3)]e^{i\lambda_{15} \alpha_{15}}.
\end{equation}
As with the $SU(2)$ subgroup ranges in $SU(3)$, the ranges for $\alpha_3$ and $\alpha_5$ 
each get multiplied by 2 and
$\alpha_{15}$'s ranges get multiplied by 4 (the number of $SU(4)$
central group elements).  The remaining ranges are either held
the same, or modified in the case of the $SU(3)$ element;
\begin{gather}
0 \le \alpha_1,\alpha_7,\alpha_{11} \le \pi,\nonumber \\
0 \le \alpha_2,\alpha_4,\alpha_6,\alpha_8,\alpha_{10},\alpha_{12}
\le \frac{\pi}{2},\nonumber \\
0 \le \alpha_3,\alpha_5,\alpha_9,\alpha_{13} \le 2\pi,\nonumber \\
0 \le \alpha_{14} \le \sqrt{3}\pi,\nonumber \\
0 \le \alpha_{15} \le 2\sqrt{\frac{2}{3}}\pi.
\end{gather}
These ranges yield both a covering of $SU(4)$, as well as the correct
group volume for $SU(4)$.

In general we can see that by looking at 
$SU(N)/Z_N$ not only can we arrive at a parametrization of $SU(N)$ 
with a logically derivable set
of ranges that gives the correct group volume, but we can also show 
how those ranges can be modified 
to cover the entire group as well without any
arbitrariness in assigning values to the parameters.
\pagebreak



\end{document}